%% file: Two_channel_Kondo-BdB.tex
\documentclass[prb,twocolumn,amsmath,amssymb,superscriptaddress,citeautoscript,longbibliography]{revtex4-2}

\usepackage{graphicx,color}
\usepackage{xcolor}
\usepackage{braket}
\usepackage{stackengine}
\usepackage{lipsum}  
\stackMath
\usepackage[colorlinks,urlcolor=blue,citecolor=blue]{hyperref}
\allowdisplaybreaks

\begin{document}

\title{Systematic compactification of the two-channel Kondo model. \\I. 
\textit{Consistent bosonization-debosonization approach and exact comparisons}}

\author{Aleksandar \surname{Ljepoja}}
\affiliation{Department of Physics, University of Cincinnati, OH-45221, USA}
\author{C.~J.~\surname{Bolech}}
\affiliation{Department of Physics, University of Cincinnati, OH-45221, USA}
\author{Nayana \surname{Shah}}
\affiliation{Department of Physics, Washington University in St.~Louis, MO-63160, USA}

\begin{abstract}

Capitalizing on recent work, that clarifies the consistent use of bosonization-debosonization methods to study Kondo-type quantum impurity models even in nonequilibrium settings, we revisit the compactification procedure of the two-channel Kondo model (by which it is rewritten more ``compactly'' using the single-channel version of the model and exploiting separation and duality between spin and charge) and uncover some hidden approximations that could limit its range of validity. This complements and extends, for two or any even number of channels, beyond previous work on the Toulouse limit of these models, and reinforces the need for the use of an extended framework in these calculations. We carry out a number of exact comparisons between the different models, and show that keeping track of the, so-called, \textit{consistency factors} leads to full agreement between the compactified and original versions of the model.
\end{abstract}

\maketitle

\section{Introduction}

In the last few decades, the investigation of non-Fermi liquids and strong correlations became one of the most active frontiers in the study of quantum materials and devices. From the theoretical standpoint, these are systems that typically require the use of creative approaches for their study since a direct application of perturbation-theory methods does not capture the full gamut of phenomena. One example of such is the separation of spin and charge into distinct branches of collective excitations. Paradigmatic models where this is observed are one-dimensional Luttinger liquids \cite{Giamarchi} and Kondo-type quantum impurities \cite{Hewson}. The latter is particularly interesting in the case of multichannel hosts, which give rise to impurity overscreening and intermediate-coupling (local) non-Fermi-liquid physics \cite{nozieres1980}.

The multichannel-Kondo fixed point, (the two-channel one in particular), is notoriously unstable to channel anisotropy; which makes its experimental realization famously difficult \cite{oreg2003,*bolech2005b,potok2007}. Moreover, some of the fixed-point signatures (crucially, the presence of a decoupled Majorana excitation) are sensitive to finite-size effects or lattice discretization. These prompted a theoretical program of finding related models that distill the main elements responsible for the non-Fermi liquid physics while, in the process, making it more accessible. (The hope was to turn channel anisotropy into a marginal perturbation, rather than a relevant one \cite{coleman1995}.) The main idea was to enforce spin-charge separation (also sensitive to finite-size and lattice effects), and use the fact that the Kondo interaction involves only the spins, of the host and the impurity, to remove the charge degrees of freedom and arrive at a more compact model (i.e., formulated in terms of a smaller number of degrees of freedom) that could be regularized without affecting the fixed-point physics that would now move from intermediate to strong coupling as in the single-channel case. This \textit{compactification} of the model was done first using symmetry and bootstrapping arguments \cite{coleman1995b} and later also via a more methodical bosonization-based mapping \cite{schofield1997}.

Bosonization relies on describing the excitations of one-dimensional fermionic systems via bosonic degrees of freedom when the dispersion is linear and no cutoff is present \cite{Gogolin,Giamarchi,vondelft1998a}. If the number of fermions in the system is held constant, the excitations over a filled Dirac-sea state are particle-hole pairs that can be decomposed in terms of bosonic operators that fully capture the corresponding spectrum. Under these conditions, bosonization is an exact correspondence. Conventionally, one of the key advantages of the bosonic reformulation is that the bosons can be easily combined to make the separation of charge and spin manifest. This is one of the enabling factors that not only makes the compactification program possible, but allows also related developments like the identification of (solvable) Toulouse points in quantum impurity models \cite{schlottmann1978,emery1992,bolech2006a,*iucci2008}.

However, even for hosts with ideal conditions for bosonization, we recently established that the presence of boundaries or impurities can interfere with exact spin-charge-flavor separation. The \textit{consistent} bosonization-debosonization (BdB) framework that we put forth correctly captures the physics of the boundary/impurity problem \cite{shah2016}. In particular, it respects the symmetries and regularizations that the \textit{conventional} BdB formalism can inadvertently violate in a wide range of problems
\footnote{In some cases, the local strong-coupling physics resulting from conventional (inconsistent) vs.~consistent BdB treatment can be so drastically different that it might misleadingly appear to be originating from a different microscopic physical setup (or ``regularization'' at the boundary \cite{[{}][{. \textit{Nota bene} that in this treatment only the fermions in one side of the junction are bosonized and there is never a recombination of bosonic degrees of freedom across the junction. We remark as well that among the different ``choices'' of interpretation of the tunneling term considered by these authors, the first one is the standard one and could correspond to a product of single-orbital fermions (microscopic fermion operators coarse-grained using the orbital wave function) that could be consistently handled via, for instance, point splitting or lattice regularization or local-action formulations; while the other ``choices'' are intrinsic bilinears and could be denoted as operator products only as a mnemonic. The first ``choice'' is therefore the relevant (implicit) one for Kondo-type impurity models that can be obtained from single-orbital Anderson-type impurities via a standard Schrieffer-Wolff transformation.}]Filippone2016}). We stress that the inconsistency addressed in Ref.~\onlinecite{shah2016} does not arise in the bosonization step itself, but rather during the subsequent manipulations of the bosonic fields. A counter-case in point is  the example in Ref.~\onlinecite{Filippone2016}, in which different ``choices'' of junctions between chiral Fermi liquids display similarly contrasting differential-conductance characteristics, but do so for clear physical reasons traceable to the microscopic differences in the models and unrelated to any emergence of conserved-quantity violations during the boson-field mixing (as in the cases we study).}.
This was first demonstrated in transport calculations across single-mode junctions and later also in a two-lead Kondo junction at the Toulouse limit \cite{bolech2016} (in qualitative agreement with subsequent numerical calculations away from the solvable point \cite{Ashida2018,*Ashida2018a}). It is thus natural to expect that the compactification of the two-channel Kondo impurity model will also be affected by the subtle breakdown of exact spin-charge-flavor separation. A systematic rederivation of the compactified model would then serve the dual purpose of exploring the limits of interaction-induced fermion fractionalization (spin-charge separation and its generalizations) while also further clarifying the essential aspects responsible for the non-Fermi-liquid physics of multichannel quantum impurities.

The purpose of this work is to present a systematic approach to the compactification of the single-impurity mutichannel (with even channel number) Kondo model that incorporates the latest understanding regarding the proper regularization of the impurity-interaction terms while performing bosonic-basis transformations. After presenting the model and setting the notation in the next section, we show in Sec.~\ref{Sec:BdB} how BdB can be used to refermionize the model into physical sectors aligned with the collective excitation of charge, spin, etc. The following section presents detailed comparisons between the original and different \textit{compact} refermionized versions of the model (by looking at three specific solvable limits). The final section closes with a general discussion that puts the different results in perspective.

\section{Two-channel Kondo Model}

We will be primarily focused on the two-channel model but generically interested in the multichannel model with an arbitrary number of channels. This is because the limit of large channel number can bring the fixed point into the weak-coupling regime (the fixed-point coupling constant scales with the inverse of the channel number). For technical reasons that will became clear in the next section, we limit ourselves to the case of even channel number and adopt a notation with two channels only (additional channels can be included at any point by introducing an extra internal index that plays no role in the subsequent transformations of the model that we carry out after bosonizing).

The case of primary interest consists thus of two channels ($L$ and $R$, when using ``leads'' notation) of noninteracting spin-$\frac{1}{2}$ fermions which are coupled via magnetic exchange to a spin-$\frac{1}{2}$ fixed-valence impurity (a spin-only degree of freedom). Explicitly, the Hamiltonian of the model is given by \cite{Cox&Z}:
\begin{equation}\label{eq:direct}
\begin{split}
    & H=H_{0}+H^{\perp}_{K}+H^{z}_{K}\\
    & H_{0} = \sum_{\sigma \ell}\int dx \ \psi^{\dagger}_{\sigma \ell}(x,t) (iv_{F}\partial_{x})\psi^{}_{\sigma \ell}(x,t) \\
    & H^{z}_{K}  =  \sum_{\sigma \ell} \sigma J_{\ell \ell}^{z} S^{z} \psi^{\dagger}_{\sigma \ell}(0,t)\psi_{\sigma\ell}(0,t)\\
    & H^{\perp}_{K} =  \sum_{\sigma,\ell} J_{\ell \ell}^{\perp} S^{\sigma} \psi^{\dagger}_{\bar{\sigma} \ell}(0,t)\psi_{\sigma\ell}(0,t)
\end{split}
\end{equation}

\noindent where we have introduced $\ell = \{L,R\}=\{-, +\}$ and $\sigma=\{\downarrow, \uparrow\} =\{-, +\}$ as lead and spin indices, respectively, and the notation $\bar{\sigma}=-\sigma$. The exchange coupling constants here are taken to be spin-anisotropic ($J_{\ell\ell}^{z}$ and $J_{\ell\ell}^{\perp}$ 
\footnote{We adopted a repeated index notation for $J^{z,\perp}_{\ell\ell}$ to match the notation of the \textit{two-lead} Kondo model, but in the (standard) \textit{two-channel} model considered here there are no $J^{z,\perp}_{LR}$ terms and a single index suffices. The case of interest most often discussed is when one has $J^{z,\perp}_{L}=J^{z,\perp}_{R}$ and the (repeated) lead/channel index can be dropped altogether.})
and factors of $1/2$ often used in the definition of the fermion spin density have been absorbed in them \cite{coleman1995,coleman1995b} (writing them explicitly is also common in the literature and amounts to a rescaling of the couplings; cf.~\cite{schofield1997,bolech2016,ye1998}). This anisotropic formulation will be natural, because the Abelian bosonization formalism that we are going to employ treats these two types of terms differently, as will be seen in the next section.

The fermionic degrees of freedom coupled to the impurity are written in terms of one-dimensional chiral fields $\psi^{}_{\sigma \ell}(x,t)$ and the energy spectrum has been linearized around the Fermi levels ($k^{\sigma\ell}_{F}=\mu^{\sigma}_{\ell}/v_{F}$). This is a well established procedure of ``unfolding'', in which one can take any impurity problem in three dimensions and write it in terms of one-dimensional fermions with a linearized spectrum. Notice that the $L$ and $R$ labels identify the channel and not the physical type of chiral mover (we use this notation because during the analysis of the models we will carry out spin transport calculations and it will be natural to refer to the two channels as left/right ``leads''). Moreover, since the impurity interaction is taken to be purely local and centered at the origin, we have the freedom to work with all left (or right) movers only, and which type we choose is a matter of convention (and one can switch from one to the other at any time via a spatial-inversion transformation that leaves the Kondo impurity terms invariant). In addition, we allow that there can be a chemical potential difference between the two fermion spin orientations (spin bias) characterized by the magnetic field of each lead, $h_{\ell} = \mu^{\uparrow}_{\ell} - \mu^{\downarrow}_{\ell}$. These fields will be important when calculating the spin current through the impurity (notice that while the standard two-channel Kondo model does not allow the flow of a charge current, it does not preclude spin transport between the two channels).
\section{Bosonization-debosonization}
\label{Sec:BdB}
Bosonization refers to a transformation that rewrites an original one-dimensional fermionic model (with a linear spectrum and no momentum cutoff \footnote{Many presentations of bosonization start by performing a \textit{low-energy approximation} in which the spectrum is first linearized, fast oscillations are factored out, and ``positron'' states added as needed; thus effectively rewriting the problem in terms of one-dimensional right- and left-moving chiral fermions \cite{Giamarchi,haldane1981}, (\textit{n.b.} there are studies of how to reincorporate the effects of band curvature, particularly in connection to nonlinear Luttinger liquids \cite{Imambekov2012}). These preliminary steps are common also to other approaches ranging from integrability \cite{andrei1983} and conformal field theory \cite{affleck1990,johannesson2003,*johannesson2005}, to the earlier implementations of the numerical renormalization group \cite{[{}][{. \textit{Corrigendum:} Although we are citing this reference in the context of the linear spectrum, ironically, Eq.~(VII.4) is missing the dispersion relation for the electrons (``$dk$'' should have read ``$dk\,k$'' in the kinetic-energy term). But the discussion in the paragraph that follows makes it clear that a linear dispersion is being used.}]Wilson1975}. Given that commonality, here we adopt the point of view that those steps are a necessary prelude to bosonization but not a part of the technique itself \cite{vondelft1998a,shah2016,bolech2016}. For the purpose of a systematic comparison, we will be using bosonization and debosonization exclusively to map between different models involving chiral fermions (for which bosonization can in principle be an ``exact mapping'' if carried out with sufficient attention to regularization and the correspondence between operators).}) 
in terms of bosonic degrees of freedom. To carry out the bosonization procedure, one needs to have a normal-ordered formulation of the problem, where the vacuum expectation value (vev) for each lead has been subtracted
\footnote{The canonical alternative for normal-ordering, from where the name originates, involves reordering of the normal-mode field operators so that the vev is automatically zero. But there is no bosonization prescription for those fermionic modes as is there for $\psi(x)$.
However, when normal-ordering continuum theories in coordinate space, as we do in the present case, point-splitting might be needed in addition to vev subtraction. So, for instance, $\rho(x)={:}\psi^\dagger(x)\psi(x){:}$ is given by
\vspace{-1mm}
\begin{equation*}
    \qquad\rho(x)\equiv\lim_{\delta \to 0} \big[ \psi^\dagger (x+\delta) \psi(x) - \langle vac \vert \psi^\dagger (x+\delta) \psi(x) \vert vac \rangle \big]
\end{equation*}
\noindent where $\vert vac \rangle$ is the vacuum (or ground state) of the theory. For further discussion, see Appendix G of Ref.~\onlinecite{vondelft1998a}, which points out when this normal-ordering prescription is not applicable and discusses the subtleties of applying it in bosonized theories}.
Notice the vev here refers to the vacuum of $H_{0}$, so it remains unmodified as interactions are switched on and off adiabatically. Although nontrivial it can be well defined even for Landauer-type nonequilibrium settings (as is the case in our model in the presence of spin bias) that we want to be able to consider. 

One way to deal with the presence of a finite bias, is to first employ a field transformation to \textit{gauge away} the chemical potentials, --bringing the model onto an effective equilibrium zero-bias setting--, and then subtract the vev for each spin orientation on each lead (see Sec.~II B of Ref.~\onlinecite{shah2016}). The price to pay is that the impurity terms acquire a time dependence. To see that, it is best to start from the model in its Lagrangian formulation,
\begin{equation}
    \begin{split}
        & L_{0} = \sum_{\sigma \ell}\int dx \ \psi^{\dagger}_{\sigma \ell}(x,t)(i\partial_{t}+iv_{F}\partial_{x})\psi^{}_{\sigma \ell}(x,t) \\
        & L^{z}_{K} =-\sum_{\sigma \ell} \sigma J_{\ell \ell}^{z} S^{z} \psi^{\dagger}_{\sigma \ell}(0,t)\psi_{\sigma\ell}(0,t)\\
        & L^{\perp}_{K} =  -\sum_{\sigma,\ell} J_{\ell \ell}^{\perp} S^{\sigma} \psi^{\dagger}_{\bar{\sigma} \ell}(0,t)\psi^{}_{\sigma \ell}(0,t)
    \end{split}
\end{equation}
Removal of the chemical potentials from the distribution functions \footnote{We chose to include the chemical potentials in the distribution functions of each lead and not to have them directly in the Lagrangian (the latter being a grand-canonical-style formulation commonly used in the Matsubara formalism, but in the presence of an applied bias it complicates the treatment of the interaction picture; see~Sec.~8.6.1 of Ref.~\onlinecite{mahan2000})} and normal ordering of the Lagrangian can now be achieved by using a gauge field transformation of the form: $\psi_{\sigma\ell}(x,t) = e^{-i( \mu^{\sigma}_{\ell}t + k^{\sigma\ell}_{F}x)} \breve{\psi}_{\sigma\ell}(x,t)$. Applying this transformation we get,
\begin{equation}
    \begin{split}
        & L_{0} = \sum_{\sigma \ell}\int dx \ : \breve{\psi}^{\dagger}_{\sigma \ell}(x,t)(i\partial_{t}+iv_{F}\partial_{x})\breve{\psi}^{}_{\sigma \ell}(x,t): \\
        & L^{z}_{K} =-\sum_{\sigma \ell} \sigma J_{\ell \ell}^{z} S^{z} :\breve{\psi}^{\dagger}_{\sigma \ell}(0,t)\breve{\psi}_{\sigma\ell}(0,t):\\
        & L^{\perp}_{K} =  -\sum_{\sigma,\ell} J_{\ell \ell}^{\perp} S^{\sigma} e^{i \bar{\sigma}h_{\ell} t} \breve{\psi}^{\dagger}_{\bar{\sigma}, \ell}(0,t)\breve{\psi}_{\sigma,\ell}(0,t)
    \end{split}
\end{equation}
It is important to note that all the information about the (spin-dependent) chemical potentials is now fully contained in the exponential factors, $e^{i \bar{\sigma}h_{\ell} t}$, in front of the spin-flip interaction term. Going back to the Hamiltonian formalism we have
\begin{equation}
    \label{eq:noH}
    \begin{split}
        & H_{0} = \sum_{\sigma \ell}\int dx \ : \breve{\psi}^{\dagger}_{\sigma \ell}(x,t)(iv_{F}\partial_{x})\breve{\psi}^{}_{\sigma \ell}(x,t): \\
        & H^{z}_{K} =\sum_{\sigma \ell} \sigma J_{\ell \ell}^{z} S^{z} :\breve{\psi}^{\dagger}_{\sigma \ell}(0,t)\breve{\psi}_{\sigma\ell}(0,t):\\
        & H^{\perp}_{K} =  \sum_{\sigma,\ell} J_{\ell \ell}^{\perp} S^{\sigma} e^{i\bar{\sigma }h_{\ell}t} \breve{\psi}^{\dagger}_{\bar{\sigma} \ell}(0,t)\breve{\psi}_{\sigma \ell}(0,t)
    \end{split}
\end{equation}
Only now, after normal ordering, can we proceed with the bosonization of each term in the Hamiltonian in the usual manner. The bosonization of the kinetic term, $H_{0}$, is standard, and we do not need to discuss it here. We shall focus on the impurity-interaction terms. As anticipated, we can see that the price for normal ordering is the introduction of an explicitly time-dependent phase in the $H^{\perp}_{K}$ term (this will be dealt with by using a suitable ``inverse'' gauge transformation after the refermionization of the model is complete). Let us discuss that term first.

\subsection{Spin-flip interaction}

The part of the Hamiltonian responsible for flipping the spin of the scattered fermion with that of the impurity is what we call the spin-flip interaction. It is given by $H^{\perp}_{K}$ in Eq.~(\ref{eq:noH}) and involves the coupling constant $J_{\ell \ell}^{\perp}$.
In order to bosonize it we use the Abelian bosonization identity (in the standard convention for left movers), 
\begin{equation}\label{Eq:Mandelstam}
\breve{\psi}_{\sigma,\ell}(x,t)=
\frac{1}{\sqrt{2 \pi a}} F_{\sigma \ell}(t)\, e^{-i\phi_{\sigma \ell}(x,t)}\,
\end{equation}
where $F_{\sigma \ell}$ are the so-called Klein factors, which ensure proper inter-species anticommutation relations; $a$ is a short-distance regularization parameter, similar but not equivalent to a lattice spacing (in particular, not introducing any interplay of charge and spin); and $\phi_{\sigma\ell}$ are new (chiral and compact) bosonic degrees of freedom. Written in terms of these, the spin-flip interaction becomes
\begin{equation}\label{eq:HperpK}
    H^{\perp}_{K} = \frac{1}{2\pi a} \sum_{\sigma,\ell} J_{\ell \ell}^{\perp} S^{\sigma}e^{i \bar{\sigma} h_{\ell}t} F^{\dagger}_{\bar{\sigma} \ell}F^{}_{\sigma \ell}e^{i\phi_{\bar{\sigma} \ell}}e^{-i\phi_{\sigma \ell}}
\end{equation}
where the bosonic fields are evaluated at $x=0$ and time $t$. 
In order to separate the description of spin ($s$) processes from those involving charge ($c$), lead ($l$) and other independent degrees of freedom ($sl$), a standard \textit{canonical} change of basis is introduced:
\begin{equation}\label{eq:ChangeBasis}
\phi_{\sigma\ell} = 
\frac{1}{2}(\phi_{c}+\sigma \phi_{s}+\ell \phi_{l}+\sigma \ell \phi_{sl})
\end{equation}
so that
\begin{equation}\label{eq:bos}
    \begin{split}
    H^{\perp}_{K} & =  \frac{\tilde{n}_{c}\tilde{n}^{-}_{l}}{2\pi a} J_{L L}^{\perp} S^{-}e^{i h_{L}t} F^{\dagger}_{\uparrow L} F_{\downarrow L} e^{i\phi_{s}}e^{-i\phi_{sl}}\\
    &\qquad +  \frac{\tilde{n}_{c}\tilde{n}^{-}_{l}}{2\pi a} J_{L L}^{\perp} S^{+} e^{-i h_{L}t} F^{\dagger}_{\downarrow L} F_{\uparrow L}e^{-i\phi_{s}}e^{i\phi_{sl}}\\
    &\qquad +  \frac{\tilde{n}_{c}\tilde{n}^{+}_{l}}{2\pi a} J_{RR}^{\perp} S^{-} e^{i h_{R}t} F^{\dagger}_{\uparrow R} F_{\downarrow R} e^{i\phi_{s}}e^{i\phi_{sl}}\\
    &\qquad +  \frac{\tilde{n}_{c}\tilde{n}^{+}_{l}}{2\pi a} J_{RR}^{\perp} S^{+} e^{-i h_{R}t} F^{\dagger}_{\downarrow R} F_{\uparrow R}e^{-i\phi_{s}}e^{-i\phi_{sl}}
    \end{split}
\end{equation}

This is where what we will call the \textit{consistent} and \textit{conventional} bosonization schemes (or frameworks) start to diverge from each other. As shown in Ref.~\onlinecite{shah2016}, where the consistent BdB formalism was introduced, the \textit{conventional} scheme amounts to having the products of vertex operators of the form $e^{\frac{i\phi_{\nu}}{2}}e^{\frac{-i\phi_{\nu}}{2}}$ [that appear after performing the change of basis in Eq.~(\ref{eq:HperpK}) and carrying out the algebra without ever mixing the two exponentials] being replaced by the identity operator (and as explained in the Appendix [p.~\pageref{Sec:Appendix}],
some gradient-of-the-field contributions are missed).
In the \textit{consistent} scheme, on the other hand, we preserve those products \textit{intact} by introducing the \textit{n-twiddle} shorthand notation for them by defining the (Fermi-density-like) operators $\tilde{n}$ as follows:
\begin{equation}\label{eq:nfactors}
    \begin{split}
        & \sqrt{2}\tilde{n}_c \equiv e^{\frac{i\phi_{c}}{2}}e^{\frac{-i\phi_{c}}{2}}\\
         & \sqrt{2}\tilde{n}^{\ell}_l \equiv e^{ \frac{i\ell \phi_{l}}{2}}e^{ \frac{-i\ell \phi_{l}}{2}}
    \end{split}
\end{equation}
We shall sometimes refer to these operators as \textit{consistency factors}. It is important to notice that, after the algebra, it would seem there is an extra overall factor of $\frac{1}{2}$ in Eq.~(\ref{eq:bos}). That factor is related to the procedure leading to \textit{consistent boundary conditions} \cite{shah2016,bolech2016}, which ensure that the $\tilde{n}$ factors behave as (square roots of) single-lattice-point densities. Understanding the origin of this factor is quite subtle and we refer the reader to Sec.~IV B of Ref.~\onlinecite{shah2016}. 

The $n$-twiddles satisfy the properties of idempotency ($(\tilde{n}^{\pm})^{2} = \tilde{n}^{\pm}$) and co-nilpotency ($\tilde{n}^{+} \tilde{n}^{-} = 0$). Their careful bookkeeping is the cornerstone of the \textit{consistent} bosonization framework. As we shall see later (and as pointed out before \cite{shah2016}; see the Appendix [p.~\pageref{Sec:Appendix}] for a brief summary), their properties are responsible for the disappearance of unphysical diagrams in different perturbative expansions. They achieve that by preventing mixing of ``normal'' and ``anomalous'' vertices in the Feynman diagrams (which we will get to later in this discussion). Note that for  $\tilde{n} \rightarrow 1$ we recover the \textit{conventional} scheme.  

One can now postulate a set of new Klein factors in the new basis (with an eye on defining new fermionic degrees of freedom) \cite{vondelft1998,zarand2000,iucci2008}. Focusing on the bosonized spin-flip interaction, we proceed further by identifying relations between bilinears of old and new Klein factors (there are four arbitrary sign choices made here, corresponding to operator-order ambiguity; we take those choices to be the same as in the previous literature),
\begin{equation}
    \begin{split}
        & F^{\dagger}_{\uparrow L} F_{\downarrow L}=F^{}_{sl} F^{\dagger}_{s}\,,\quad
        F^{\dagger}_{\downarrow L} F_{\uparrow L}=-F^{\dagger}_{sl} F^{}_{s}\\
        & F^{\dagger}_{\uparrow R} F_{\downarrow R}=F^{\dagger}_{sl} F^{\dagger}_{s}\,,\quad
        F^{\dagger}_{\downarrow R} F_{\uparrow R}=-F^{}_{sl} F^{}_{s}
    \end{split}
\end{equation}
With these relationships between bilinears, we arrive at the final form for the bozonized spin-flip interaction in the new basis,
\begin{equation}
    \begin{split}
    H_{K}^{\perp} & =\frac{\tilde{n}_{c}\tilde{n}^{-}_{l}}{2\pi a} J_{LL}^{\perp} S^{-} e^{i h_{L}t} F^{}_{sl}F^{\dagger}_{s} e^{i\phi_{s}}e^{-i\phi_{sl}} \\
    & \qquad -\frac{\tilde{n}_{c}\tilde{n}^{-}_{l}}{2\pi a} J_{LL}^{\perp} S^{+} e^{-i h_{L}t} F^{\dagger}_{sl}F^{}_{s}e^{-i\phi_{s}}e^{i\phi_{sl}}\\
    & \qquad +\frac{\tilde{n}_{c}\tilde{n}^{+}_{l}}{2\pi a} J_{RR}^{\perp} S^{-}e^{i h_{R}t} F^{\dagger}_{sl}F^{\dagger}_{s} e^{i\phi_{s}}e^{i\phi_{sl}} \\
    & \qquad -\frac{\tilde{n}_{c}\tilde{n}^{+}_{l}}{2\pi a} J_{RR}^{\perp} S^{+}e^{-i h_{R}t} F^{}_{sl}F^{}_{s} e^{-i\phi_{s}}e^{-i\phi_{sl}}
    \end{split}
\end{equation}
Using the bosonization identity to define a set of \textit{physical-basis} fermions:
\begin{equation}
\breve{\psi}_{\nu}(x,t)=
\frac{1}{\sqrt{2 \pi a}} F_{\nu}(t) e^{-i\phi_{\nu}(x,t)}
\end{equation}
(with $\nu = c,s,l,sl$), we can now debosonize the spin-flip interaction in terms of these fermions that are aligned with the different physical processes and conservation laws. Explicitly,
\begin{equation}
    \begin{split}
        H_{K}^{\perp} & = \tilde{n}_{c}\tilde{n}^{-}_{l}J_{L L}^{\perp} \Big(S^{-} e^{i h_{L}t} \breve{\psi}_{sl}(0,t)\breve{\psi}^{\dagger}_{s}(0,t)\\
        & \qquad \qquad \qquad  - S^{+} e^{-i h_{L}t} \breve{\psi}^{\dagger}_{sl}(0,t)\breve{\psi}_{s}(0,t) \Big)\\
        &+  \tilde{n}_{c}\tilde{n}^{+}_{l}J_{RR}^{\perp} \Big( S^{-} e^{i h_{R}t} \breve{\psi}_{sl}^{\dagger}(0,t)\breve{\psi}_{s}^{\dagger}(0,t)\\
        & \qquad \qquad \qquad  - S^{+} e^{-i h_{R}t} \breve{\psi}_{sl}^{}(0,t)\breve{\psi}_{s}^{}(0,t) \Big)\\
    \end{split}
\end{equation}
Proceeding further with the refermionization of the model, we need to deal with the explicit time dependence in the exponential factors. It is the inverse problem from the one we tackled before the bozonization, where we translated from a finite to a zero bias setting and normal ordered it. In order to gauge away the explicit time dependence from $H^{\perp}_{K}$, we choose the chemical potentials of $s$ and $sl$ sectors to be $\mu_{s}=\frac{1}{2} (h_{R}+h_{L})$ and $\mu_{sl}=\frac{1}{2} (h_{R}-h_{L})$. (Notice this choice is not unique but all other choices can be seen to be physically equivalent, and normal order does not play a role in the case of the spin-flip interaction.) Therefore, upon carrying out a gauge transformation of the form $\psi_{\nu}(x,t) = e^{-i \mu_{\nu}(t + \frac{x}{v_{F}})} \breve{\psi}_{\nu}(x,t)$, we finally recover a refermionized spin-flip interaction without explicit time dependence:
\begin{equation}\label{eq::refermionized}
    \begin{split}
        H_{K}^{\perp} & = \tilde{n}_{c}\tilde{n}^{-}_{l}J_{L L}^{\perp} \left(S^{-} \psi_{sl}(0)\psi^{\dagger}_{s}(0)- S^{+} \psi^{\dagger}_{sl}(0)\psi_{s}(0) \right)\\
        &+ \tilde{n}_{c}\tilde{n}^{+}_{l}J_{RR}^{\perp} \left( S^{-} \psi_{sl}^{\dagger}(0)\psi_{s}^{\dagger}(0) - S^{+} \psi_{sl}^{}(0)\psi_{s}^{}(0) \right)\\
    \end{split}
\end{equation}
As we can see from this expression, there are two terms in the refermionized spin-flip part of the Hamiltonian. One is associated with $\tilde{n}^{-}_{l}$ and $J^{\perp}_{LL}$, and it is the one that contains ``normal'' vertices. The other one comes with $\tilde{n}^{+}_{l}$ and $J^{\perp}_{RR}$, and contains what we call ``anomalous'' vertices (products of two creation or two annihilation operators). In the \textit{conventional} scheme, since there are no $\tilde{n}$ factors (in the expression above, they can be set to one, $\tilde{n}\to 1$), diagrams that mix the anomalous and normal vertices are not prohibited. It is these diagrams that give unphysical contributions in any perturbative calculation and are the sure cause of discrepancies between different BdB schemes. We shall elaborate more on this in the coming sections, but now we switch to discussing the non-spin-flip part of the Kondo interaction.

\subsection{Parallel interaction}
The part of the Hamiltonian in which electrons scatter on the impurity without changing their spin is what we call the parallel (or non-spin-flip) interaction, and it is given by
\begin{equation}
    \begin{split}
        H^{z}_{K} =\;:\!H^{z}_{K}\!:\;= 
        \sum_{\sigma \ell} \sigma J_{\ell \ell}^{z} S^{z}\,:\!\breve{\psi}^{\dagger}_{\sigma \ell}(0,t)\breve{\psi}_{\sigma\ell}(0,t)\!:
    \end{split}
\end{equation}
For this interaction term, the difference between how bosonization is carried out in each scheme -- and the resulting refermionized Hamiltonians -- is more striking than in the case of the spin-flip interaction term. So it is best to first present each BdB scheme separately and later compare the final results. 

In the case of the \textit{conventional} bosonization scheme one uses the standard bosonization identity, and performs a point-splitting procedure, to arrive at
\begin{equation}\label{eq:norder}
\begin{split}
:\!\psi^{\dagger}_{\sigma \ell} (x,t) \psi^{}_{\sigma \ell}(x,t)\!:\,&=
\psi^{\dagger}_{\sigma \ell} (x,t) \psi^{}_{\sigma \ell}(x,t)-\frac{1}{2}\\
& = \frac{1}{2\pi} \partial_{x} \phi_{\sigma \ell}(x,t)
\end{split}
\end{equation}
(the fermion vev is assumed to be $1/2$ for any species and at any point in space \cite{schofield1997} in the ideal-host situation required for ``exact'' bosonization) and gets
\begin{equation}
    \begin{split}
        H^{z}_{K} & =\frac{1}{2\pi}\sum_{\sigma, \ell} \sigma J_{\ell \ell}^{z} S^{z}\partial_{x}\phi_{\sigma \ell}(0,t)\\
        & = \frac{1}{2\pi}(J^{z}_{LL} + J^{z}_{RR}) S^{z}\partial_{x}\phi_{s}(0,t) \\ 
        & \qquad - \frac{1}{2\pi}(J^{z}_{LL} - J^{z}_{RR}) S^{z}\partial_{x}\phi_{sl}(0,t)
    \end{split}
\end{equation}
where, in the second equality, the same change of basis as before was used
\footnote{To compare with Ref.~\cite{bolech2016}, one needs to notice that the Hamiltonians are normalized differently, namely, $J_\mathrm{here}=J_\mathrm{there}/2$. After accounting for that, there is still an extraneous factor of 1/2 in Eq.~(17) which needs fixing but has no other implications for the results of that paper. Then the Toulouse limit corresponds to $J^{z}_\mathrm{there}\to2\pi v_F$, which here is $J^{z}_\mathrm{avg}\to\pi v_F$}. 
Debosonizing the parallel interaction, with a similar identity used in reverse, one arrives at the following expression:
\begin{equation}\label{eq:parallelC}
    \begin{split}
        H^{z}_{K} 
        & = (J_{RR}^{z}+J_{LL}^{z})\,S^{z}\,:\!\psi^{\dagger}_{s}(0,t)\psi^{}_{s}(0,t)\!:\\
        & \quad + (J_{RR}^{z}-J_{LL}^{z})\,S^{z}\,:\!\psi^{\dagger}_{sl}(0,t)\psi^{}_{sl}(0,t)\!:
    \end{split}
\end{equation}
On the other hand, carrying out the bosonization in the \textit{consistent} manner, we refrain from using the bosonization identity for the diagonal fermion bilinears that was just used in the conventional scheme. The reason is that such an identity assumes that the limit of $a\to0$ has already been taken (yielding the derivative of the bosons), and it does not allow to later carry out a change of basis without assuming that the product of vertex operators of opposite signs is \textit{one}. Instead, in the \textit{consistent} scheme, we start from the expression for the parallel interaction, and we proceed to bosonize it using the standard identity for individual fermionic fields that we used for bosonizing the spin-flip part of the interaction. Upon doing this transformation, and a subsequent rotation into the physical-sector basis, we find that the bosonized non-spin-flip interaction is given by
\begin{equation} \label{eq:hzbos}
    \begin{split}
    H^{z}_{K} & = \frac{2}{a \pi}S^{z}\tilde{n}_{c}\tilde{n}^{-}_{l}J_{LL}^{z} \big ( \tilde{n}^{+}_{s}\tilde{n}^{-}_{sl} - \tilde{n}^{-}_{s}\tilde{n}^{+}_{sl} \big ) \\
    & \quad + \frac{2}{a \pi}S^{z}\tilde{n}_{c} \tilde{n}^{+}_{l} J_{RR}^{z} \big ( \tilde{n}^{+}_{s}\tilde{n}^{+}_{sl} - \tilde{n}^{-}_{s}\tilde{n}^{-}_{sl} \big )\\
    & = \frac{2}{a \pi}S^{z}\tilde{n}_{c}\tilde{n}^{-}_{l}J_{LL}^{z} \big ( \tilde{n}^{+}_{s} - \tilde{n}^{+}_{sl} \big ) \\
    & \quad + \frac{2}{a \pi}S^{z}\tilde{n}_{c} \tilde{n}^{+}_{l} J_{RR}^{z} \big ( \tilde{n}^{+}_{s} - \tilde{n}^{-}_{sl} \big )
    \end{split}
\end{equation}
where we have introduced the $\tilde{n}$ notation as defined in Eq.~(\ref{eq:nfactors}), and in the second equality we have used properties of the $\tilde{n}$ factors to simplify the expression (namely, $\tilde{n}^{+}_{\nu}+\tilde{n}^{-}_{\nu} = 1$, to the desired order in an operator product expansion). We know that $\tilde{n}_{l}$ and $\tilde{n}_{c}$ are constants of motion, because there is no overall change of total fermion number, nor are there any fermions going from the $L$ to the $R$ lead (i.e., there are no inter-lead terms), in the original formulation of the two-channel Kondo model. Therefore, we keep those two $\tilde{n}$ while we choose to rewrite the remaining ones (those in the $s$ and $sl$ sectors) using the identity $\frac{1}{\pi a}\tilde{n}_{\nu} =  \psi^{\dagger}_{\nu} \psi^{}_{\nu}$, which we get from combining idempotency and the squaring of both sides in Eq.~(\ref{eq:nfactors}),
\begin{equation}
          2 \tilde{n} = 2 \tilde{n}^2 = e^{i\phi}e^{-i\phi} = 2 \pi a \psi^{\dagger} \psi\\
\end{equation}
Using this identity and, in the second step, normal ordering as outlined in Eq.~(\ref{eq:norder}), we arrive at

\begin{equation}
    \begin{split}
    H^{z}_{K} 
    & = S^{z}\tilde{n}_{c}\tilde{n}^{-}_{l}J_{LL}^{z} \big ( \psi^{\dagger}_{s}\psi^{}_{s} - \psi^{\dagger}_{sl}\psi^{}_{sl} \big ) \\
    & \;\; + S^{z}\tilde{n}_{c} \tilde{n}^{+}_{l} J_{RR}^{z} \big ( \psi^{\dagger}_{s}\psi^{}_{s} - (1-\psi^{\dagger}_{sl}\psi_{sl}) \big ) \\
    & = S^{z}\tilde{n}_{c}\tilde{n}^{-}_{l}J_{LL}^{z} \big ( : \! \psi^{\dagger}_{s}\psi^{}_{s} \! : - : \! \psi^{\dagger}_{sl}\psi^{}_{sl}\! : \big ) \\
    & \;\; + S^{z}\tilde{n}_{c} \tilde{n}^{+}_{l} J_{RR}^{z} \big ( : \! \psi^{\dagger}_{s}\psi^{}_{s}\! : + : \! \psi^{\dagger}_{sl}\psi_{sl}\! : \big ) \\
    & = \tilde{n}_{c}(\tilde{n}^{+}_{l} J_{RR}^{z}+\tilde{n}^{-}_{l}J_{LL}^{z})\,S^{z}
    : \! \psi^{\dagger}_{s}(0,t) \psi^{}_{s}(0,t) \! : \\
    & \;\; + \tilde{n}_{c} (\tilde{n}^{+}_{l} J_{RR}^{z}-\tilde{n}^{-}_{l}J_{LL}^{z})\,S^{z}
    : \! \psi^{\dagger}_{sl}(0,t)\psi_{sl}(0,t)\! :
    \end{split}
\end{equation}

This is the BdB-refermionized parallel interaction in the \textit{consistent} scheme. In a similar way as in the case of the spin-flip interaction, the parallel interaction has two parts: one with the $\tilde{n}^{-}_{l}$ factor and the $L$-lead coupling constant, and the other with the $\tilde{n}^{+}_{l}$ factor and the $R$-lead coupling constant. Setting $\tilde{n}^{\pm} \to 1$ we recover the same expression as in the \textit{conventional} scheme in Eq.~(\ref{eq:parallelC}) in a similar way as is the case with the perpendicular (spin-flip) interaction. Notice also that in the channel-symmetric case, with $J_{RR}^{z}=J_{LL}^{z}$, the consistency factors drop out from the spin sector of the model, which is the same in both schemes (up to factors of $2$ due to varying boundary conditions \cite{shah2016}); but the spin-lead sector does not ``disappear'' in the \textit{consistent} case due to the presence of the $\tilde{n}$-difference operator that is generically nonzero (and responsible for an alternating sign in transport-setting calculations).

\section{Comparing conventional and consistent $\mathrm{\mathbf{BdB}}$-compactifications against direct calculations} \label{sec::compare}

Over the next subsections, we are going to look at three different simple limits of the two-channel Kondo model that are not generic but allow for exact calculations: (a) the flat-band, (b) the $x$-axis-Ising, and (c) the Toulouse limits. In the first of these limits, the system is finite, and we are going to calculate the (spin-sector) electron's Green function and self-energy. We will compare them as obtained from the \textit{conventional} and \textit{consistent} BdB-refermionized versions of the model, and contrast them also with those calculated within the original formulation of the model. On the other hand, in the other two limits, the system is infinite and can be modeled in out-of-equilibrium situations. We are going to focus on the spin current that is driven by a finite magnetic-field gradient between channels (i.e., across the impurity) and compare the results between before and after BdB refermionization (in both schemes).

\subsection{Flat-band Limit}
\label{sec:flat-band}
One limit of the two-channel Kondo model for which we can calculate interesting physical quantities in both the \textit{direct} and BdB-compactified formulations of the model is the so-called flat-band (or narrow-band
\footnote{This nomenclature is of extended use in the recent literature on topological materials. Sometimes, in the literature on other types of systems, ``flat-band'' is used to refer to a constant density of states per energy (and often considered in the wide-band limit). That is a different situation than the one we study in here and it should not be confused}) limit \cite{alascio1979,*alascio1980,*proetto1981,Tran2018,*Kumar2021,*Checkelsky2024}.
In this limit, electrons in the leads have a single energy level available to them with energy that we are going to set to zero. (It can also be regarded as the limit of zero Fermi velocity of linear-dispersion channels with the band reduced to a single state per channel for each spin.) In this simple yet very instructive limit, the Hilbert space is finite and relatively small. Thus the Hamiltonians are amenable to be exactly diagonalized. 

The \textit{direct} flat-band Hamiltonian is given by
\begin{equation}\label{eq:fbD}
    \begin{split}
        H^{\perp}_{K}= & J_{\perp} S^{+} \sum_{\ell}c_{\ell, \downarrow}^{\dagger} c^{}_{\ell, \uparrow}+J_{\perp}S^{-} \sum_{\ell} c_{\ell, \uparrow}^{\dagger} c_{\ell,\downarrow}\\
        H^{z}_{K}= & J_{z} S^{z} \sum_{\ell} (c_{\ell,\uparrow}^{\dagger} c^{}_{\ell, \uparrow}-c_{\ell, \downarrow}^{\dagger} c^{}_{\ell,\downarrow})
    \end{split}
\end{equation}
where $\ell$ counts the two channels ($\ell = \{L,R\} = \{-1,+1\}$). The kinetic part of the Hamiltonian is identically zero, because of the limit, or we can also consider a generic value of the flat-band energy level and later set it to $\varepsilon^{}_{\ell,\sigma} \rightarrow 0$. 

On the other hand, the flat-band limit of the BdB-refermionized Hamiltonian is given by (the kinetic term is again zero)
\begin{equation}\label{eq:fbC}
    \begin{split}
        H^{\perp}_{K}= & 
        - J_{\perp}\tilde{n}_{c}\tilde{n}^{-}_{l}
        \left( S^{+} c_{sl}^{\dagger} c^{}_{s} 
        + S^{-} c_{s}^{\dagger} c_{sl} \right) \\
        & \quad 
        - J_{\perp}\tilde{n}_{c}\tilde{n}^{+}_{l} 
        \left( S^{+}  c_{sl}^{} c^{}_{s} 
        + S^{-}  c_{s}^{\dagger} c^{\dagger}_{sl} \right) \\
        H^{z}_{K}= & 
        J_{z} \tilde{n}_{c}\tilde{n}^{-}_{l} S^{z} 
        \left ( c_{s}^{\dagger} c^{}_{s}
        - c_{sl}^{\dagger} c^{}_{sl} \right ) \\
        & \quad + J_{z} \tilde{n}_{c}\tilde{n}^{+}_{l} S^{z}  
        \left ( c_{s}^{\dagger} c^{}_{s}
        + c_{sl}^{\dagger} c^{}_{sl} -1 \right )
    \end{split}
\end{equation}
As we have seen in previous sections, for a specific replacement value of the $\tilde{n}$ factors (namely, $\tilde{n} \to 1$) one recovers the \textit{conventionally} refermionized Hamiltonian. Alternatively, preserving the key operator properties of the $\tilde{n}$ factors is what we call the \textit{consistently} refermionized model. 

Both direct and (conventionally) refermionized Hamiltonians can be exactly diagonalized, which is an important characteristic of the flat-band limit. By means of a full diagonalization of the Hamiltonian matrix, one easily finds its eigenstates and eigenvalues, as well as evaluates its partition function (either numerically or even symbolically, using a computer-algebra system). In that way one can explore the full spectrum and the exactly obtainable thermodynamic properties of the model.

However, because this is a very specific limit of a (reduced) bulk given by a single level with its energy taken to be zero, one \textit{a priori} does not necessarily expect the bosonization mappings to have matching limits. This is because in the constructive-bosonization procedure one has to take the limits of zero Fermi velocity (for a ``narrow band'') and infinite bandwidth consistently, and there could be problems with the order of limits.
That would be the naive explanation for any mismatch we observe when comparing thermodynamic quantities between the \textit{conventional} and \textit{direct} models, since even in the direct model alone (no BdB involved) some results change if one takes the zero-Fermi-velocity limit first or at the end of the calculation. This could be seen more clearly in the diagrammatic calculations for generalized flat-band models with arbitrary $\varepsilon^{}_{\ell,\sigma}$ that we shall discuss later \cite{Ljepoja2024c}. (Incidentally, the results from the exact diagonalization of the flat-band limit will then be useful to double-check the results of the frequency summations involved in the evaluation of self-energy diagrams.)

We readily find that the partition functions differ between the direct and conventionally refermionized models. As a reflection of it, the usual construction for the ``residual impurity entropy'' in the \textit{conventional} case gives, in units of the Boltzmann constant, $S^{}_\mathrm{imp}\!=\!\ln(1/2)$. Meanwhile, the \textit{direct} model gives $S_\mathrm{imp}\!=\!\ln(1/8)$ in the flat-band limit \footnote{However, it would have been $S^{}_\mathrm{imp}\!=\!\ln(\sqrt{2})$ for any nonzero Fermi velocity (\textit{i.e.}, taking the zero-temperature limit first). This illustrates how the order of limits matters, as pointed out in the discussion. The residual impurity entropy is known to be very sensitive in this respect, (cf.~with $S^{}_\mathrm{imp}\!=\!\ln(2)$ in the case of the finite-size zero-temperature limit for any non-zero Fermi velocity \cite{zarand2000})}. This is an interesting result because in the BbB procedure we refermionized the bulk (electronic) degrees of freedom, while keeping the impurity unchanged, so it might seem surprising that there is such a difference when taking the flat-band limit in the two cases. But one of the reasons for the difference is in the different eigenvalue degeneracies that arise in this limit after reorganizing the degrees of freedom in the Hilbert space during the BdB mapping. It is thus not surprising that the ground-state degeneracies, and thus the residual entropies, are affected.

On the other hand, calculating the residual impurity entropy in the \textit{consistent} scheme (handling the $\tilde{n}$ carefully, as will be detailed later in this section) we recover the direct-calculation result. This would suggest that even in this very specific limit
the \textit{consistent} way of compactifying gives results that are in line with the direct calculation. More than this agreement, however, it is the discrepancy between the limits of the two compactification schemes (despite being defined in identical Hilbert spaces) that indicates a logical need of further comparisons. It is also relevant to note that at high temperatures all the schemes coincide and give the expected value of $S_\mathrm{imp}\!=\!\ln(2)$, since, in that limit of large thermal fluctuations, the impurity is effectively decoupled from the bulk and has two available states (the spin-up and spin-down orientations) regardless of the scheme.

Having compared the schemes in the context of thermodynamic quantities, we next want to focus on the dynamic properties; --that depend not only on the eigenvalues, but also on the eigenvectors. Specifically, we choose to compare the self-energies of the $s$-sector electrons between the limits of the two BdB-refermionized models, (\textit{conventional} and \textit{consistent}). We explicitly use the Lehman representation of the finite-temperature Green function in the Matsubara formalism,
\begin{equation}
    \begin{split}
        G_{s}(i \omega_{n}) = \frac{1}{Z}\sum_{n,m} \left|\bra{n} c_{s} \ket{m}\right|^{2} \;\frac{e^{-\beta E_{n}}+e^{-\beta E_{m}}}{i \omega + E_{n} - E_{m}}
    \end{split}
\end{equation}
where $\ket{n}$ and $E_{n}$ are the eigenstates and corresponding eigenvalues of the Hamiltonian, respectively. Knowing the full Matsubara Green function, one can extract the self-energy using the standard Dyson equation. For the \textit{conventional} scheme the calculation is straightforward: $\tilde{n}$ factors disappear from the Hamiltonian (allowing for the unphysical mixing of different vertices in the perturbative expansion), and we can find the Green functions and self-energy by simple diagonalization. 
\begin{figure}[t]
    \centering
    \includegraphics[width=\columnwidth]{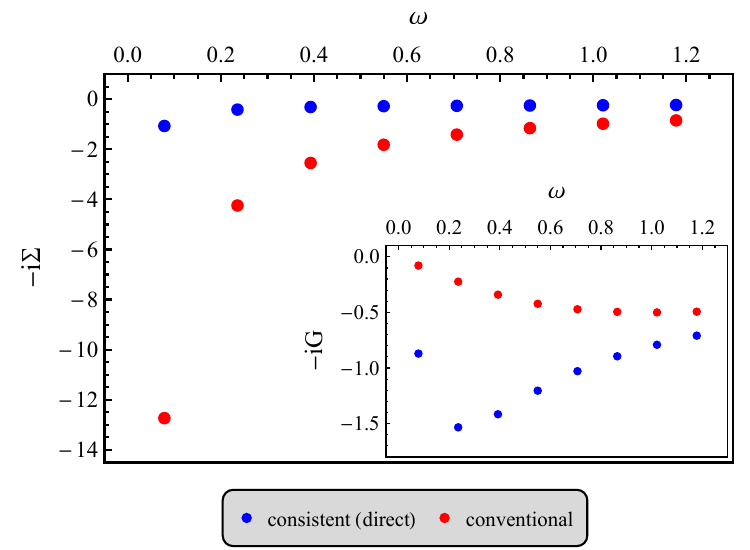}
    \caption{Behavior of the imaginary part of the fermion self-energy ($-i \Sigma_{}$) as a function of Matsubara frequency. The inset shows the Green function ($-iG_{}$) as a function of frequency. Both plots are for a specific value of the isotropic-case coupling constant, $J=0.5$, and for $\beta = 40$. One can see that, for small frequencies, the \textit{conventional} and \textit{consistent} schemes disagree to a significant degree.}
    \label{fig:omegadep}
\end{figure}
On the other hand, the \textit{consistent} case needs a more careful treatment. In order to find the self-energy, one needs to find a way to carefully deal with the $\tilde{n}^{+}_{l}$ and $\tilde{n}^{-}_{l}$ factors that appear in the Hamiltonian in Eq.~(\ref{eq:fbC}). One possible way to take care of those factors is to define them as actual densities of particles in the $l$ sector (mathematically, this amounts to choosing a lattice-like regularization and picking a particular branch of the square root), namely, replacing $\tilde{n}^{+}_{l} \to c_{l}^{\dagger} c^{}_{l}$ and $\tilde{n}^{-}_{l} \to c_{l}^{} c^{\dagger}_{l}$. That way, any combination of the anomalous and normal vertices will carry a product of the $l$-sector particle and hole densities, which is zero by construction. Therefore, all of the unphysical diagrams are now excluded. However, this way of doing calculations is over-restrictive; meaning that some diagrams that are physically allowed are nevertheless also excluded by this method. Those diagrams are the fourth- and higher-order ones in the expansion for the self energy that have fermion loops in them. On physical grounds, a fermion loop in such diagrams can involve, say, anomalous vertices while the rest of the diagram has normal ones, or \textit{vice-versa}. This would be a physically allowed contribution, but the method where the $\tilde{n}$ factors are defined as densities would not include it, because it involves products of hole and particle densities. As a consequence, by underestimating the multiplicity factor of diagrams with fermion loops, this naive method fails to capture the full multichannel character of the original model (but notice it becomes evident only at high orders in perturbation theory).

\begin{figure}[t]
    \centering
    \includegraphics[width=\columnwidth]{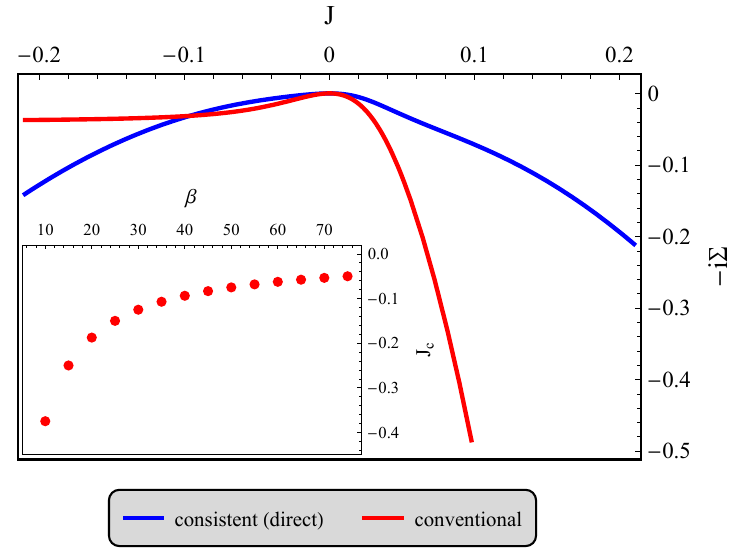}
    \caption{Comparison of the fermion self-energies between different BdB schemes. The self-energy is plotted as a function of the coupling constant $J_{\perp} = J_{z} \equiv J$ for a fixed frequency $\omega=\frac{\pi}{\beta}$, and $\beta = 40$. The inset shows the value of coupling constant ($J_{c}$) at which the \textit{conventional} self-energy crosses the direct one, as a function of temperature.}
    \label{fig:SEcomparison}
\end{figure}

Alternatively, one can cast the \textit{consistent} model into an effective-two-channel model where one channel corresponds to $\tilde{n}^{-}_{l}$ while the other one corresponds to $\tilde{n}^{+}_{l}$. Introducing a channel index to the fermions that matches the $n$-twiddle index of each term and then taking $\tilde{n}\to1$ we have
\begin{equation}\label{eq:fbCS}
    \begin{split}
        H^{\perp}_{K}= & - J_{\perp}S^{+} c_{sl,-}^{\dagger} c^{}_{s,-} - J_{\perp}S^{-}  c_{s,-}^{\dagger} c_{sl,-} \\
        & \qquad -J_{\perp} S^{+}  c_{sl,+}^{} c^{}_{s,+} - J_{\perp}S^{-}  c_{s,+}^{\dagger} c^{\dagger}_{sl,+}\\
        H^{z}_{K}= & J_{z} S^{z} \big ( c_{s,-}^{\dagger} c^{}_{s,-}-c_{sl,-}^{\dagger} c^{}_{sl,-} \big ) \\
        & \qquad +  J_{z} S^{z}  \big ( c_{s,+}^{\dagger} c^{}_{s,+}+c_{sl,+}^{\dagger} c^{}_{sl,+} -1 \big )\\
    \end{split}
\end{equation}
Because of the independence of these channels, any mixing of the anomalous and normal vertices is now prevented, but in such a way that fermionic loops can be seen as corresponding to \textit{independent n-twiddle histories}. This effective-two-channel \textit{consistent} Hamiltonian is still not quadratic, but it is possible to exactly diagonalize it and find the Green function and self-energy in the same way as for the \textit{conventional} model. The results of such calculations are displayed in Fig.~\ref{fig:omegadep}, where we show the behavior of the $s$-sector self-energy ($\Sigma_{s}$) as a function of Matsubara frequency. In the inset we show the behavior of the (purely imaginary) Green function as well. 
By studying these plots, one can see that the disagreement between the \textit{conventional} and \textit{consistent} schemes is greatest in the low-frequency regime. In particular, for the lowest Matsubara frequency ($\omega = \frac{\pi}{\beta}$) there is an order of magnitude difference between the \textit{conventional} and \textit{consistent} results. To further illustrate the difference between the schemes, in Fig.~\ref{fig:SEcomparison} we show the self-energy as a function of the coupling constant for the lowest Matsubara frequency, (for which we have observed the largest disagreement between the two). Even for small values of the coupling constant disagreement between the \textit{conventional} and the \textit{consistent} calculations is clearly noticeable, and only in the limit of $J\to0$ do the results match. Moreover, one can see that for negative values of the coupling constant the asymptotic large-frequency behaviors are strikingly different. While the \textit{conventional} self-energy saturates to a value of $-\frac{1}{2} \omega$, that is not the case for the \textit{consistent} calculation. In the inset of Fig.~\ref{fig:SEcomparison} we show the values of the coupling constant ($J_{c}$) at which the \textit{conventional} self-energy crosses the \textit{consistent} result, and starts saturating, as a function of the inverse temperature. From the behavior of $J_{c}$ one can observe that at a low temperatures the \textit{conventional} scheme breaks down (and its self-energy saturates) ``sooner'' than at a high temperatures; which points in the direction of a modified infrared fixed point.

The source of all these listed discrepancies should be looked for in the ``mixing" of the regular and anomalous vertices in the diagrams of the perturbative expansion of the self-energy. Such mixings correspond to unphysical processes (that would be violating channel conservation if expressed in the direct formulation of the model) and generate the disagreement between the \textit{consistent} and \textit{conventional} schemes. 
 Furthermore, inspecting the Hamiltonian in Eq.~(\ref{eq:fbCS}), we can see that it is isomorphic to the direct-model Hamiltonian upon the transformation
 \begin{equation}
    \label{eq:transformation}
     \begin{split}
        & c^{\dagger}_{s,-} \to c^{\dagger}_{L,\uparrow} \qquad \quad \ c^{\dagger}_{s,+} \to c^{\dagger}_{R,\uparrow} \\
        & c^{\dagger}_{sl,-} \to - \ c^{\dagger}_{L,\downarrow} \qquad c^{}_{sl,+} \to - \ c^{\dagger}_{R,\downarrow}
     \end{split}
 \end{equation}
Because the two models are connected by such a mapping, we expect that the corresponding self-energies for each degree of freedom match between the \textit{consistent} and \textit{direct} formulations. Indeed, that is what we observe.
Stepping back, this connection constitutes a compelling indication in favor of the superiority of the \textit{consistent} BdB-refermionization scheme over the \textit{conventional} one. Since it would always produce matching results with the direct calculation even when considering a delicate scenario like the flat-band limit.

\subsection{$x$-axis Ising limit}

A limit of the anisotropic Kondo model in which the spin-exchange interaction is only along a single (easy) axis is called an Ising limit. We will choose it to be the $x$-axis. In this limit the impurity does not have any dynamic (its $x$-axis projection is a conserved quantity), and the only remaining degrees of freedom in the interaction term are from the electrons. The problem is then Gaussian, and therefore amenable to a trivial exact solution, both directly and also after compactification. In terms of the original fermions, the Hamiltonian of the $x$-axis Ising limit is obtained by setting $J^{y}_{\ell \ell} = J^{z}_{\ell \ell} =0$ in an XYZ-exchange version of Eq.~(\ref{eq:direct}), which gives
\begin{eqnarray}
    H &=&H_{0}+H^{x}_{K}\notag\\
    H_{0} &=& \sum_{\sigma \ell}\int dx \ \psi^{\dagger}_{\sigma \ell}(x,t) (iv_{F}\partial_{x})\psi^{}_{\sigma \ell}(x,t) \\
    H^{x}_{K} &=&  \sum_{\sigma,\ell} J_{\ell \ell}^{x} S^{x} \psi^{\dagger}_{\bar{\sigma} \ell}(0,t)\psi^{}_{\sigma\ell}(0,t)\notag
\end{eqnarray}

As can be seen from the Hamiltonian, the interaction entails only spin-flip processes (traced over spin). That means that the only way electrons scatter off from the impurity is by changing their spin. Our model is set up in such a way that we have a spin bias on each lead 
\footnote{In this subsection, as well as in the next one where we study spin transport at the Toulouse point, the two-channels are taken to be \textit{leads} with spin-dependent chemical potentials. Notice that we are still considering the standard two-channel Kondo model, (\textit{i.e., without} $J_{LR}$ charge co-tunneling terms added to the Hamiltonian).} 
measured by the lead magnetic fields $h_{L}$ and $h_{R}$, which are given as the difference between spin-dependent chemical potentials that enter directly into the distribution functions for each lead in a Landauer-type formulation. We shall not assume anything about the relative values of these fields and keep the calculation as general as possible. These spin biases are responsible for driving a spin current (and they would correspond to standard potential biases and charge currents, respectively, in the case of a ``charge Kondo circuit,'' which has received renewed theoretical and experimental attention recently \cite{matveev1991,*matveev1995,*Furusaki1995,Iftikhar2015,*Iftikhar2018,Karki2022,Pouse2023}). 

This limit of the Kondo model, being exactly solvable, lends itself as a good testing ground to compare the different BdB-compactification schemes and the direct calculation among each other. The spin-current operator is defined as the change of the spin difference between the two leads over time, and it is given by
\begin{widetext}
\begin{equation}
        I_{S}=\partial_{t} \frac{\Delta S}{2} =\frac{i}{2} [H,\Delta S ]=\frac{i}{2} [H^{x}_{K},\Delta S ]
        =\frac{i}{4}\sum_{\sigma \ell} J^{x}_{\ell \ell} S^{x} \ell (\sigma -\bar{\sigma})\psi^{\dagger}_{\bar{\sigma} \ell} \psi^{}_{\sigma\ell} \\        
\end{equation}
Where $\Delta S$ is the spin difference across leads, given by $\Delta S=\frac{1}{2}\sum_{\sigma, \ell} \sigma \ell \psi^{\dagger}_{\sigma \ell}\psi^{}_{\sigma \ell}$. Since there are spin biases, our model is out of equilibrium (we assume a simplified Landauer-style setup with spin-dependent chemical potentials \cite{katsura2007}) and in order to calculate the average value of said current we turn to the Keldysh formalism and choose the following spinor basis:
\begin{equation}
        \Psi = \left(\psi^{\kappa=-}_{\uparrow L} \quad
        \psi^{\kappa=+}_{\uparrow L} \quad 
        \psi^{\kappa=-}_{\uparrow R} \quad 
        \psi^{\kappa=+}_{\uparrow R} \quad 
        \psi^{\kappa=-}_{\downarrow L} \quad 
        \psi^{\kappa=+}_{\downarrow L} \quad 
        \psi^{\kappa=-}_{\downarrow R} \quad 
        \psi^{\kappa=+}_{\downarrow R}\right)^{T}
\end{equation}
\newline
where the new index $\kappa$ labels the Keldysh branch following a ``minus-means-forward'' convention. In this basis, the inverse local Green function at the impurity site (the regularized local action, cf.~Ref.~\onlinecite{shah2016}; see Ref.~\onlinecite{bolech2004,*bolech2005,*bolech2007,*kakashvili2008a} for other examples) is given by the matrix

\begin{equation}
    \begin{split}
        \frac{G^{-1}(\omega)}{2 v_{F}} =
        \begin{pmatrix}
            is_{\uparrow L}&-is_{\uparrow L}+i&0&0&-2S^{x}t_{L}&0&0&0\\
            -is_{\uparrow L}-i&is_{\uparrow L}&0&0&0&2S^{x}t_{L}&0&0\\
            0&0&i s_{\uparrow R}&-i s_{\uparrow R}+i&0&0&-2S^{x}t_{R}&0\\
            0&0&-i s_{\uparrow R}-i&is_{\uparrow R}&0&0&0&2S^{x}t_{R}\\
            -2S^{x}t_{L}&0&0&0&is_{\downarrow L}&-is_{\downarrow L}+i&0&0\\
            0&2S^{x}t_{L}&0&0&-is_{\downarrow L}-i&is_{\downarrow L}&0&0\\
            0&0&-2S^{x}t_{R}&0&0&0&is_{\downarrow L}&-is_{\downarrow L}+i\\
            0&0&0&2S^{x}t_{R}&0&0&-is_{\downarrow L}-i&is_{\downarrow L}\\
        \end{pmatrix}
    \end{split}
\end{equation}
\end{widetext}
where we introduced the notations $J^{x}_{\ell \ell} = 4 v_{F} t_{\ell}$ and $s_{\sigma \ell} = \tanh(\frac{\omega-\mu^{\sigma}_{\ell}}{T_{\ell}})$, and $T_{\ell}$ is the temperature of each lead. Having $G^{-1}(\omega)$, we can invert it to find the appropriate Green functions in the Keldysh formalism and use them in the expression for the spin current (the average expectation value of the spin-current operator):
\begin{equation}\label{eq:currentopD}
    \begin{split}
        \langle I_{S} \rangle &=\frac{1}{4}\sum_{\sigma \ell} J^{x}_{\ell \ell} S^{x} \ell (\sigma -\bar{\sigma}) \int \frac{d \omega}{2 \pi} G^{-+}_{\sigma\ell, \bar{\sigma} \ell}(\omega) \\
    \end{split}
\end{equation}
Replacing with the explicit expressions we get
\begin{equation}
    \begin{split}
        \langle I_{S} \rangle &=\int^{\infty}_{-\infty} \frac{d \omega}{2\pi} \frac{\big ( s_{\downarrow L}(\omega)-s_{\uparrow L}(\omega)\big )t_{L}^2}{1+t^{2}_{L}} \\
        & \qquad \qquad + \int^{\infty}_{-\infty} \frac{d \omega}{2\pi}\frac{\big ( -s_{\downarrow R}(\omega)+s_{\uparrow R}(\omega) \big ) t^{2}_{R}}{1+t^{2}_{R}}\\
    \end{split}
\end{equation}
Although the $\omega$ integrals can be calculated for any temperature, in order to keep the expressions simple we turn to the zero-temperature limit. In this limit the $s_{\sigma\ell}(\omega)$ functions become Heaviside step functions and the integrals are trivial to evaluate:
\begin{equation}\label{eq:currentXD}
    \begin{split}
        \langle I_{S} \rangle &=  \frac{h_{L}t^{2}_{L}}{\pi (1+t^{2}_{L})^2}-\frac{h_{R}t^{2}_{R}}{\pi (1+t^{2}_{R})^2}
    \end{split}
\end{equation}
We can see that the total spin current consists of the sum of the spin current leaving the $L$ lead and that entering the $R$ lead, and the two are independent. The two current components are driven by the respective lead magnetic fields, and for an isotropic model and the special configuration of equal fields, $h_{L}=h_{R}$, the current is identically zero. This is to be expected since, for that configuration of the magnetic fields, the difference of the lead spins stays the same over time due to the fact that on both leads identical demagnetization processes are taking place, (which flip spin up to spin down fermions at the same rate).

Having the expression for the spin current in the direct formulation of the model, we can turn our attention to the \textit{conventional} scheme and compare with the  expression we get in that case. In the \textit{conventional} BdB scheme the x-Ising limit is compactly refermionized into
\begin{equation}
    \begin{split}
        H_{K}^{x} & = J_{L L}^{x} S^{x} \bigg(\psi_{sl}(0)\psi^{\dagger}_{s}(0)- \psi^{\dagger}_{sl}(0)\psi_{s}(0) \bigg)\\
        & \qquad + J_{RR}^{x} S^{x} \bigg(\psi_{sl}^{\dagger}(0)\psi_{s}^{\dagger}(0) - \psi_{sl}^{}(0)\psi_{s}^{}(0) \bigg)\\
    \end{split}
\end{equation}
The spin current, as was the case in the direct calculation, is given by the change of $\Delta S$ over time, where $\Delta S$ is the spin difference between the two leads. In the refermionized language, because the basis is rotated into the physical sectors, the spin difference is given by $\Delta S = N_{sl}$. The spin current operator is then given as the commutator of $N_{sl}$ and the refermionized Hamiltonian:
\begin{widetext}
\begin{equation} \label{eq:currentCB}
        I_{S}=\partial_{t} \frac{\Delta S}{2} =\frac{i}{2} [H^{x}_{K},N_{sl} ]
        =\frac{i}{2} J^{x}_{LL} S^{x}  \bigg(\psi_{sl} \psi^{\dagger}_{s}+  \psi^{\dagger}_{sl} \psi_{s} \bigg) 
         - \frac{i}{2} J_{RR}^{x} S^{x} \bigg(\psi_{sl}^{\dagger}\psi_{s}^{\dagger} - \psi_{sl}^{}\psi_{s}^{} \bigg)\\
\end{equation}
In order to get the average spin current, the appropriate Green functions need to be obtained. We can calculate them proceeding similarly as before but with with the necessary generalizations. We are still adding a Keldysh index to deal with the nonequilibrium character of the problem, but this time we need to introduce a Nambu structure as well, due to presence of ``anomalous'' processes. We adopt the following spinor basis:
\begin{equation}
    \begin{split}
        \Psi = \left( \psi^{-}_{s}(\omega) 
               \quad \psi^{+}_{s}(\omega) 
               \quad \psi^{- \dagger}_{s}(\bar{\omega}) 
               \quad \psi^{+ \dagger}_{s}(\bar{\omega}) 
               \quad \psi^{-}_{sl} (\omega) 
               \quad \psi^{+}_{sl}(\omega) 
               \quad \psi^{- \dagger}_{sl}(\bar{\omega}) 
               \quad \psi^{+ \dagger}_{sl}(\bar{\omega}) \right)^{T}
    \end{split}
\end{equation}
where $\bar{\omega}=-\omega$ and we restricted the frequencies to the positive semiaxis, (in order to avoid double counting). The inverse Green function matrix is given by
\begin{equation}\label{eq:localC}
    \begin{split}
        \frac{G^{-1}(\omega)}{2 v_{F}} =
        \begin{pmatrix}
            is_{s}&-is_{s}+i&0&0&2S^{x}t_{L}&0&2S^{x}t_{R}&0\\
            -i s_{s}-i&i s_{s}&0&0&0&-2S^{x}t_{L}&0&-2S^{x}t_{R}\\
            0&0&i\bar{s}_{s}&-i\bar{s}_{s}+i&-2S^{x}t_{R}&0&-2S^{x}t_{L}&0\\
            0&0&-i\bar{s}_{s}-i&i\bar{s}_{s}&0&2S^{x}t_{R}&0&2S^{x}t_{L}\\
            2S^{x}t_{L}&0&-2S^{x}t_{R}&0&is_{sl}&-is_{sl}+i&0&0\\
            0&-2S^{x}t_{L}&0&2S^{x}t_{R}&-is_{sl}-i&is_{sl}&0&0\\
            2S^{x}t_{R}&0&-2S^{x}t_{L}&0&0&0&i\bar{s}_{sl}&-i\bar{s}_{sl}+i\\
            0&-2S^{x}t_{R}&0&2S^{x}t_{L}&0&0&-i\bar{s}_{sl}-i&i\bar{s}_{sl}\\
        \end{pmatrix}
    \end{split}
\end{equation}
\end{widetext}
where, similarly as before, we defined: $s_{\nu} = \tanh(\frac{\omega-\mu_{\nu}}{2 T_{\nu}})$, and $\bar{s}_{\nu} = \tanh(\frac{\omega+\mu_{\nu}}{2 T_{\nu}})$. Getting the appropriate Green functions, by inverting $G^{-1}(\omega)$, and using them in Eq.~(\ref{eq:currentCB}), we find that the average spin current is now given by
\begin{equation}
    \begin{split}
        \langle I_{S} \rangle &=\int^{\infty}_{0} \frac{d \omega}{2 \pi} \frac{\big ( s_{s}(\omega)-\bar{s}_{s}(\omega)\big ) \big(t_{R}^2 - t_{L}^2 \big)}{t^{4}_{L}-2t^{2}_{L} \big( t^{2}_{R} - 1 \big)+\big( t^{2}_{R} + 1 \big)^2} \\
        & \qquad  + \int^{\infty}_{0} \frac{d \omega}{2 \pi}\frac{\big ( s_{sl}(\omega)-\bar{s}_{sl}(\omega)\big ) \big(t_{R}^2 + t_{L}^2 \big)}{t^{4}_{L}-2t^{2}_{L} \big( t^{2}_{R} - 1 \big)+\big( t^{2}_{R} + 1 \big)^2}\\
    \end{split}
\end{equation}

Again this time, the integrals in $\omega$ can be done for finite temperature, but for the sake of simplicity we restrict ourselves to the zero-temperature limit. Doing them we find the average spin current in the \textit{conventional} scheme to be:
\begin{equation}\label{eq:currentXC}
    \begin{split}
        \langle I_{S} \rangle &=-\frac{1}{\pi}\;
        \frac{\mu_{s} \big(t_{R}^2 - t_{L}^2 \big)+ \mu_{sl} \big(t_{R}^2 + t_{L}^2 \big)}
        {\big(t_{R}^2 - t_{L}^2 \big)^2 + 2\big(t^{2}_{R} + t^{2}_{L}\big) + 1}\\
        &=\frac{1}{\pi}\;
        \frac{ h_{L}t_{L}^2 - h_{R}t_{R}^2}
        {\big(t_{R}^2 - t_{L}^2 \big)^2 + 2\big(t^{2}_{R} + t^{2}_{L}\big) + 1}
    \end{split}
\end{equation}
Where, in going to the second line, we have used the fact that $\mu_{s}=\frac{1}{2} (h_{R}+h_{L})$ and $\mu_{sl}=\frac{1}{2} (h_{R}-h_{L})$. Comparing the expression for the spin current in Eq.~(\ref{eq:currentXC}) with the one obtained in the direct calculation, given by Eq.~(\ref{eq:currentXD}), we see that in the \textit{conventional} scheme there is a ``mixing'' of the processes from the $L$ and $R$ leads. Instead of the total spin current being the difference of the demagnetization currents produced separately in each lead, we get a more complicated expression as evidenced by the combined appearance of $t_{R}$ and $t_{L}$ in the common denominator. Notice that if either of $t_{R}$ or $t_{L}$ is zero, the respective lead is decoupled from the impurity (its spin is conserved) and the spin current reduces to the demagnetization current of the coupled lead, --with the exact same expression as found in the direct calculation, Eq.~(\ref{eq:currentXD}).

\begin{figure}
    \centering
    \includegraphics[width=\columnwidth]{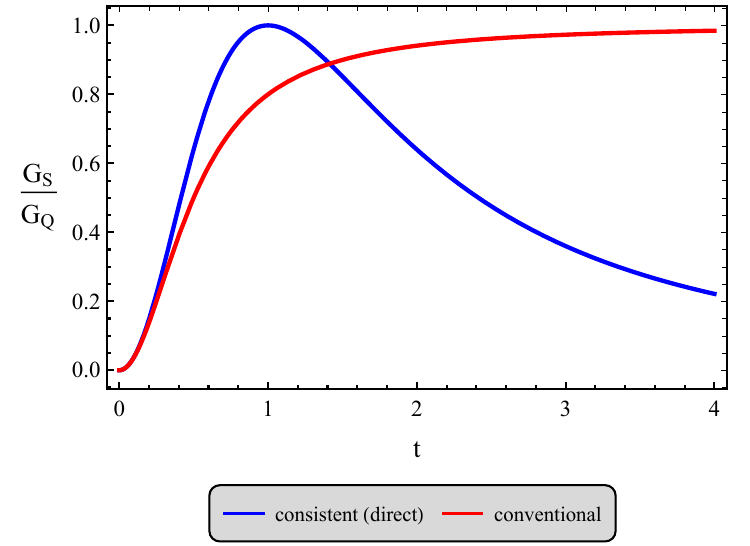}
    \caption{Comparison of differential spin conductance, ($G_S=dI_S/dh$ with $h=h_{L}=-h_{R}$), calculated exactly in both the \textit{conventional} and the \textit{consistent} schemes, the latter coinciding with the direct calculation as explained in the text. The plot is in units of the quantum of conductance, $G_{Q}=\frac{1}{2\pi \hbar}$, and as a function of the isotropic coupling constant $t=t_{L}=t_{R}$.}
    \label{fig:conductance1}
\end{figure}

A further comparison of the \textit{conventional} and direct schemes is given in Fig.~\ref{fig:conductance1} for the symmetric model and anti-aligned lead magnetic fields ($h=h_{L}=-h_{R}$). We can see that, for a small coupling constant $t$, the differential spin conductances, though not matching, exhibit similar behavior (as $t \rightarrow 0$ they in fact match). This agreement is evident from the power expansions, where in the second order in $t$ the \textit{conventional} and direct results match but for higher orders they do not. This is as expected, because unphysical diagrams (the ones mixing the ``anomalous'' and ``normal'' vertices; cf.~\cite{shah2016}) are expected to appear at fourth and higher orders in a perturbative expansion. In addition, for $t \rightarrow \infty$, the \textit{conventional}-scheme result for the differential conductance saturates to the conductance quantum, while in the direct calculation it reaches that value (at $t=1$) but then goes back down to zero (reminiscent of the similarly contrasting results for the charge current in the two-lead Kondo model \cite{schiller1998a,bolech2016}).

The unphysical mixing of the ``normal'' and ``anomalous'' vertices disappears in the \textit{consistent}-scheme calculation due to properties of the $\tilde{n}$ factors. In that case the Hamiltonian is given by
\begin{equation}
    \begin{split}
        H_{K}^{x} & = \tilde{n}_{c}\tilde{n}^{-}_{l}J_{L L}^{x}S^{x} \bigg( \psi_{sl}(0)\psi^{\dagger}_{s}(0)- \psi^{\dagger}_{sl}(0)\psi_{s}(0) \bigg)\\
        &+ \tilde{n}_{c}\tilde{n}^{+}_{l}J_{RR}^{x} S^{x} \bigg(  \psi_{sl}^{\dagger}(0)\psi_{s}^{\dagger}(0) - \psi_{sl}^{}(0)\psi_{s}^{}(0) \bigg)\\
    \end{split}
\end{equation}
The local-action matrix is exactly the same as in the conventional case in Eq.~(\ref{eq:localC}), except that this time the coupling constant $t_{L}$ is multiplied by $\tilde{n}_{l}^{-}$ and the coupling constant $t_{R}$ is multiplied by $\tilde{n}^{+}_{l}$ (the overall factor of $\tilde{n}_{c}$ is inconsequential for this discussion) and because of the co-nilpotence property ($\tilde{n}^{+}_{l} \tilde{n}^{-}_{l} = 0$) any processes that would have mixed the $L$ and $R$ leads are zero. One way of calculating the spin current is by calculating it separately for the two eigenvalues of $\tilde{n}_{l}$. We can first take $\tilde{n}_{l}^{-}=1$ and $\tilde{n}_{l}^{+}=0$ and calculate the spin current (basically obtaining the contribution from the $L$ lead). Afterwards, we can repeat the calculation for the other case, with $\tilde{n}_{l}^{-}=0$ and $\tilde{n}_{l}^{+}=1$, obtaining the contribution from the $R$ lead. Adding the two we get the total spin current to be exactly as in Eq.~(\ref{eq:currentXD}). And, of course, the results match at finite temperatures as well.

\subsection{Toulouse Point}

By \textit{Toulouse point} (or \textit{Toulouse line}) one refers to an exactly solvable limit in the parameter space of the one- or two-channel Kondo or Anderson single-impurity model in which the Hamiltonian is quadratic in terms of refermionized degrees of freedom \cite{schlottmann1978,emery1992,kotliar1996,bolech2006a,*iucci2008}. The transformation of the two-channel Kondo model into its Toulouse point is achieved by the following unitary transformation \cite{emery1992}
\begin{equation}
    U = e^{i \gamma_{s} \phi_{s}S^{z}}
\end{equation}
which is applied after bosonization of the model and rotation into physical sectors, as outlined in previous sections, and followed by the subsequent debosonization to recover a Hamiltonian given in terms of fermionic fields (for a full derivation, see Sec. III of Ref.~\onlinecite{bolech2016} and consider the special case of $J^{\perp}_{LR} = J^{\perp}_{RL} = 0$). This transformation mixes band and impurity degrees of freedom and goes beyond the simple compactification of the model. Going into the Toulouse limit we set $\gamma_{s} = 1$, which in the two-channel case is sometimes called the Emery-Kivelson transformation. After such a transformation, the parallel Kondo interaction is fine-tuned so as to be absorbed into the kinetic term and we are left only with the spin-flipping term that, after the transformation, acquires the following bosonized form
 \begin{eqnarray}
    H_{K}^{\perp} =&\;\; \frac{\tilde{n}_{c}\tilde{n}^{-}_{l}}{2 \pi a}
    J_{LL}^{\perp}  e^{i h_{L}t} S^{-}F^{}_{sl}F^{\dagger}_{s}e^{-i\phi_{sl}} \notag\\
    & -\frac{\tilde{n}_{c}\tilde{n}^{-}_{l}}{2 \pi a}
    J_{LL}^{\perp}  e^{-i h_{L}t} S^{+}F^{\dagger}_{sl}F^{}_{s}e^{i\phi_{sl}}\notag\\
    & +\frac{\tilde{n}_{c}\tilde{n}^{+}_{l}}{2 \pi a}
    J_{RR}^{\perp}e^{i h_{R}t}  S^{-}F^{\dagger}_{sl}F^{\dagger}_{s} e^{i\phi_{sl}} \\
    & -\frac{\tilde{n}_{c}\tilde{n}^{+}_{l}}{2 \pi a}
    J_{RR}^{\perp} e^{-i h_{R}t} S^{+}F^{}_{sl}F^{}_{s}e^{-i\phi_{sl}}\notag
\end{eqnarray}
Further, one can define $d^{\dagger}=S^{+}F_{s}$ \cite{zarand2000}, which (upon a similar debosonization as those discussed previously) leads to a refermionized spin-flipping interaction of the form
\begin{eqnarray}\label{eq:tolouseH}
        H_{K}^{\perp} = & \tilde{n}_{c}\tilde{n}^{-}_{l}J_{L L}^{\perp} 
        \left( \psi_{sl}(0)d^{}_{}- \psi^{\dagger}_{sl}(0) d^{\dagger} \right)\notag\\
        &+ \tilde{n}_{c}\tilde{n}^{+}_{l}J_{RR}^{\perp} 
        \left( \psi_{sl}^{\dagger}(0)d^{}_{} - \psi_{sl}^{}(0)d^{\dagger}_{} \right)
\end{eqnarray}
where we have chosen the resonant-level and spin-lead-sector chemical potentials to be $\mu_{d}=-\frac{1}{2} (h_{R}+h_{L})$ and $\mu_{sl}=\frac{1}{2} (h_{R}-h_{L})$, respectively, in order to gauge away from the Hamiltonian the time-dependent phases (after all the BdB transformations were already completed). 

One can calculate spin currents in a similar way as we did above for the case of the x-Ising limit. However, since the Toulouse limit corresponds to a strong-coupling limit of the original model, we do not have any exact direct calculation to compare results with. One can only compare \textit{conventional} and \textit{consistent} schemes with each other, and one can still look at which of them corresponds to the expected physical picture of spin transport at the Toulouse point. We are going to look at the \textit{conventional} BdB scheme first. It is obtained by letting $\tilde{n}_{c} \rightarrow 1$ and $\tilde{n}_{l} \rightarrow 1$ in the Hamiltonian of Eq.~(\ref{eq:tolouseH}). The resulting Hamiltonian is now quadratic despite corresponding to a strong-coupling situation in terms of the original degrees of freedom.

To calculate we shall be using again the Keldysh local-action method to address the question of spin transport, (including as well a Nambu representation of the degrees of freedom due to the presence of ``anomalous'' vertices in the model). To take all that into account, we adopt the following spinor basis that mirrors the one used in the previous section:
\begin{widetext}
\begin{equation}
    \begin{split}
        \Psi = \bigg( d^{-}_{}(\omega) 
        \quad d^{+}_{}(\omega) 
        \quad d^{- \dagger}_{}(\bar{\omega}) 
        \quad d^{+ \dagger}_{}(\bar{\omega}) 
        \quad \psi^{-}_{sl} (\omega) 
        \quad \psi^{+}_{sl}(\omega) 
        \quad \psi^{- \dagger}_{sl}(\bar{\omega}) 
        \quad \psi^{+ \dagger}_{sl}(\bar{\omega}) \bigg )^{T}
    \end{split}
\end{equation}
(with the $s$-band fermion now replaced by the impurity $d$ fermion that absorbed it after the rotation). Defining, as before, $s_{\nu} = \tanh(\frac{\omega-\mu_{\nu}}{2 T_{\nu}})$ and $\bar{s}_{\nu} = \tanh(\frac{\omega+\mu_{\nu}}{2 T_{\nu}})$, we get the local-action matrix:
\begin{equation}\label{eq:localC1}
    \begin{split}
        \frac{G^{-1}(\omega)}{2 v_{F}} =
        \begin{pmatrix}
            \omega-\mu_{d}&0&0&0&-t_{RR}&0&-t_{LL}&0\\
            0&-\omega+\mu_{d}&0&0&0&t_{RR}&0&t_{LL}\\
            0&0&\omega+\mu_{d}&0&t_{LL}&0&t_{RR}&0\\
            0&0&0&-\omega-\mu_{d}&0&-t_{LL}&0&-t_{RR}\\
            -t_{RR}&0&t_{LL}&0&is_{sl}&-is_{sl}+i&0&0\\
            0&t_{RR}&0&-t_{LL}&-is_{sl}-i&is_{sl}&0&0\\
            -t_{LL}&0&t_{RR}&0&0&0&i\bar{s}_{sl}&-i\bar{s}_{sl}+i\\
            0&t_{LL}&0&-t_{RR}&0&0&-i\bar{s}_{sl}-i&i\bar{s}_{sl}\\
        \end{pmatrix}
    \end{split}
\end{equation}
where we have used the rescaled coupling constants $J^{\perp}_{LL}=2v_{F}t_{LL}$ and $J^{\perp}_{RR}=2v_{F}t_{RR}$ (notice the factor-of-two difference with the similar rescaling used in the previous section). The expression for the average spin current can be found in a similar way as we did for the $x$-Ising-limit calculation. Namely, by evaluating the commutator of the $\Delta S = N_{sl}$ operator and the interaction Hamiltonian. This gives
\begin{equation} \label{eq:currentT}
        \langle I_{S} \rangle =\frac{i}{2} J^{\perp}_{LL} \left( \langle \psi_{sl} d^{}_{}\rangle -   \langle d^{\dagger}_{} \psi^{\dagger}_{sl}\rangle  \right) - \frac{i}{2} J_{RR}^{\perp} \left( \langle \psi_{sl}^{\dagger}d_{}^{}\rangle - \langle d^{\dagger}_{} \psi_{sl}^{}\rangle \right)
\end{equation}
Finding the appropriate Green-function matrix elements, one can calculate the spin current
\small
\begin{equation}\label{eq:currentC}
\begin{split}
\langle I_{S} \rangle = \int_{0}^{\infty} \frac{d \omega}{2 \pi}\frac{8 \ \omega^2 \ t^{2}_{LL} \  t^{2}_{RR}  \ (s_{sl}(\omega) - \bar{s}_{sl}(\omega))}{t^{8}_{LL}-4 \ t^{6}_{LL} \ t^{2}_{RR}+t^{8}_{RR} + (\omega^2-\mu^{2}_{d}) + 2 \ t^{4}_{RR} \ (\omega^2+\mu^{2}_{d})+4 \ t^{2}_{LL} \ t^{2}_{RR} \ (3 \ \omega^2 - t^{4}_{RR} - \mu^{2}_{d})+2 \  t^{4}_{LL} \ (\omega^2 + 3t^{4}_{RR}+\mu^{2}_{d})}
\end{split} 
\end{equation}
\normalsize
\end{widetext}
The integral in $\omega$ in the expression for current is rather complicated but it might be illustrative to do it for the special case when the magnetic fields of the two leads are equal in magnitude but of opposite alignment ($h_{L} = -h_{R} \equiv h $, and  $\mu_{sl} = -h$) and in the spin isotropic limit ($t_{LL}= t_{RR} \equiv t/2$). In that combined limit, the current becomes
\begin{equation}
\begin{split}
\langle I_{S} \rangle & = \frac{1}{2\pi} \int_{0}^{\infty} d \omega \ \frac{ t^{4}}{2 (t^4+\omega^2)} (s_{sl}(\omega) - \bar{s}_{sl}(\omega)) \\
& \stackunder{\longrightarrow}{T\to0} \; \frac{1}{2\pi} \int_{0}^{-\mu_{sl}} d \omega \ \frac{ t^{4}}{(t^4+\omega^2)} \\
& \qquad \qquad \qquad \qquad \qquad
= \frac{t^2}{4 \pi} \arctan \bigg ( \frac{h}{t^2} \bigg )
\end{split} 
\end{equation}
In doing the integral in $\omega$ we took the zero-temperature limit so that $s_{sl}(\omega) - \bar{s}_{sl}(\omega)$ becomes a Heaviside step function. Having the expression for the average current we can find the differential conductance in the same way we did for the $x$-axis Ising limit, (by taking the derivative with respect to the lead magnetic field). In the case of zero bias ($h=0$), we get that the differential conductance is one half of the quantum of conductance, and it is independent of the strength of the coupling constant $t$.

Having obtained the \textit{conventional} result, we can now look at the \textit{consistent} scheme. Since the only difference between these schemes is the now explicit presence of the $\tilde{n}$ factors, all the formal calculation of the spin current remains the same. We will have the same spinor and the same local-action matrix, with the only difference of a redefinition of the coupling constants, where $t_{LL}$ will carry a factor of $\tilde{n}^{-}_{l}$ while $t_{RR}$ will have a factor of $\tilde{n}_{l}^{+}$. Studying the full integral expression for the spin current given above, we can see that the numerator contains a product of the two coupling constants associated with the $L$ and $R$ leads. Thus, due to the co-nilpotence property of the consistency factors, this product is always zero in the \textit{consistent} scheme and, consequently, the spin current is identically zero at all temperatures. Looking in terms of differential conductance we see that, for the zero bias configuration, the \textit{conventional} scheme gives a constant differential conductance of $\frac{1}{2}G_{Q}$ while the \textit{consistent} schemes gives a zero differential conductance, --both regardless of the coupling strength.

Since we do not have an exact \textit{direct} result to compare with and determine which scheme yields the correct answer, we can instead resort to physical-plausibility arguments to decide it. 
For that, let us refer to the schematic depiction given in Fig.~\ref{fig:2chK} and discuss first the situation with a single channel (say $L$, for concreteness). Since the lead is magnetized with more up than down spins, there can be an excess of up spins at the location of the impurity (whose spin will tend to point down due to the antiferromagnetic coupling). Along the Toulouse line, the $J_{z}$ coupling is constant and \textit{maximal} (as compared to $J_{\perp}$
\footnote{With our conventions, the Toulouse limit corresponds to $J_{z}=\pi v_{F}$ for the two-channel case (or $2\pi v_{F}$ for single-channel). These coincide with the RG fixed-point values $g_{z,\perp}^{\star}=2\rho_{0}J_{z,\perp}^{\star}=1$ (or $2$, respectively) as obtained from the universal terms of the beta function \cite{Ljepoja2024b}. As the temperature (or the cutoff) is lowered, $J_{z}$ will not flow, since it is already at its maximal (\textit{i.e.}, fixed-point) value, while $J_{\perp}$ will flow to the fixed-point value (and restore spin isotropy) starting from weak coupling; cf.~Ref.~\cite{johannesson2003,*johannesson2005}. (Note that for the single-channel case non-perturbative arguments indicate that the fixed point corresponds to \textit{infinite} coupling \cite{coleman1995,nozieres1974}.) It is in that context that we can use $J_{\perp}\ll J_{z}$ to reason physically about the transport processes along the Toulouse line. This and similar kinds of considerations are also often presented in terms of phase shifts and \textit{unitary-limit} scattering \cite{schofield1997,kotliar1996,fabrizio1995a,*fabrizio1995b,Cox&Z}.}) ,
and antialignment is an energetically favorable configuration; but a spin-flip process could let a spin from the lead flip down while the impurity flips up without any change of the energy due to the $J_{z}$ coupling. The resulting high-energy down-spin electron in the lead could relax due to intra-lead processes implicit in the Landauer setup, but in the absence of other leads the impurity is locked pointing up and the process will not repeat (and even if it did, the lead magnetization would fluctuate in opposite directions each time, since the total spin of the system is conserved). 

If we now include a second channel with opposite magnetization, it will allow for the opposite spin-flip processes at the impurity. For relatively small values of $J_{z}$, the impurity is now not locked and can alternatively flip up and down while interacting with the $L$ and $R$ channels, respectively. This allows a continuous demagnetization process to take place in locked step at both leads. However, as the $J_{z}$ coupling grows stronger, an initially down-pointing impurity would attract an up spin from the $R$ lead (despite the opposite magnetization of the lead; due to which it will not have a tendency to flip the impurity). If now a $L$-lead up-spin electron tries to flip the impurity, the process is \textit{frustrated}, since it is not exchange-energy neutral any longer; because flipping the impurity would make it ferromagnetically aligned with the $R$ electron (an energetically unfavorable configuration). A large parallel interaction is thus expected to suppress the demagnetization process of the leads and eventually make the spin current go to zero, as also predicted by the \textit{consistent} Toulouse-point calculation. On the other hand, the \textit{conventional} calculation seems to correspond to a situation when $J_{z}$ and $J_{\perp}$ are always comparable. In the absence of the $n$-twiddle factors, different channels do not have separate dynamics and the frustration of alternate spin-flip processes is lost.

Beyond two channels, there is no Toulouse limit \cite{tsvelik1995}, but one can still try to address the case of \textit{maximal} parallel interaction for the multi-two-channel version of the model (notice one would still be able to find a Toulouse-like point if one were to neglect the Klein factors). In this case, the larger the number of channels the more effective would the parallel interaction be in blocking spin-flip processes; because the fixed-point scaling of $J_{z}$ is compensated by the sum of the channel contributions, but not so for the scaling of $J_{\perp}$, leading to a suppressed impurity dynamics. 

\section{Summary and Discussion}

To summarize the insights gained from our work and to frame the discussion, let us rewrite the Hamiltonian of Eq.~(\ref{eq:direct}) explicitly for the spin-isotropic case,
\begin{equation}\label{eq:directiso}
\begin{split}
    & H=H_{0}+H_{K}\\
    & H_{K} =  \sum_{\ell,\alpha} J_{\ell \ell}\;\vec{\sigma}_{\ell\alpha}(0) \cdot \vec{S}
\end{split}
\end{equation}
where $\sigma^{i}_{\ell\alpha}(x)=\psi^{\dagger}_{\sigma'\ell\alpha}(x) \, [\sigma^{i}]_{\sigma'\sigma''} \, \psi^{}_{\sigma''\ell\alpha}(x)$ is the (doubled) spin density of the $\ell\alpha$-th channel; --the sums over repeated indices are implicit. Here $[\sigma^{i}]$ are the usual Pauli matrices, and we introduced an additional internal index to the fermions ($\alpha=1,\dots,M$) so that the (even) total number of channels is $2M$.

After a systematic BdB-refermionization (separately for each value of $\alpha$), the model can be recast as \footnote{We will be including a minus sign in the definition of the second component of the spinor, [see below and cf.~with Eq.~(\ref{eq:transformation})], but we could have alternatively absorbed that minus sign in a redefinition of the perpendicular coupling before specializing to the spin-isotropic case.}
\begin{equation}\label{eq::refermionizediso}
    \begin{split}
        & H=H_{0}+H_{K}\\
        & H_{K} =  \sum_{\ell,\alpha} \tilde{n}_{c, \alpha}\tilde{n}^{\ell}_{l, \alpha}J_{\ell\ell}\;\vec{\tau}_{\ell\alpha}(0) \cdot \vec{S}
    \end{split}
\end{equation}
with $\vec{\tau}_{\ell\alpha}(x) = \Psi^{\dagger}_{\ell\alpha}(x)\cdot[\vec{\sigma}]\cdot\Psi_{\ell\alpha}(x)$, where the spinor is 
defined as $\Psi_{L\alpha}=(\psi_{s,\alpha}\;\,-\!\psi_{sl,\alpha})^{T}$ for $\ell\!=\!L$, or as $\Psi_{R\alpha}=(\psi_{s,\alpha}\;\,-\!\psi^{\dagger}_{sl,\alpha})^{T}$ for $\ell\!=\!R$, respectively. If one now follows the \textit{conventional} scheme of approximating $\tilde{n}\to1$, then the $c$ and $l$ sectors are completely decoupled from the impurity and one recovers the usual \textit{compactified} version of the Kondo model \cite{coleman1995,coleman1995b} that is specified with only half as many band degrees of freedom as compared with the \textit{direct} formulation in Eq.~(\ref{eq:directiso}).

In contrast with the single-channel Kondo model, the case of two channels is special (and generalizes to any even number of channels). Because while in both cases one observes the phenomena of spin-charge separation, --unveiled by the bosonization of the models--, the additional two-valued channel index allows for a further fractionalization and reorganization of the degrees of freedom that leads to a natural debosonization. In turn, this systematic BdB-refermionization shows that the separation of charge, spin, and channel is not complete. Specifically, a subtle coupling of the channel sector to the spin sector, that manifests itself at the \textit{boundary} (i.e., the location of the impurity), is responsible for capturing the full relation between spin and channel conservation during the spin-exchange processes with the impurity. The \textit{conventional} approximation can then be interpreted as neglecting the correlation effects of this residual coupling and enabling the complete compactification of the model. Interestingly, the two-channel case can be ubiquitous, arising from the two chiralities present when considering impurities in Luttinger liquids, multi-impurity models, or their combinations \cite{Gogolin,kane1992a,*Zachar1996,*Zhang2017,*Bortolin2022}. An additional side effect of the residual coupling of different sectors is manifest in the nondecoupling of emergent Majorana degrees of freedom at the impurity site; --they decouple only in the \textit{conventional} scheme \cite{emery1992,Lopes2020,*Komijani2020}. The implications of this for the prospects of (topological) quantum-information storage deserve to be studied further.

The comparisons presented in Sec.~\ref{sec::compare} demonstrate that, for $M\!=\!1$, a \textit{consistent} treatment of Eq.~(\ref{eq::refermionizediso}), or its spin-anisotropic extension, recovers the exact same results as in the original model, while a \textit{conventional} ``mean-field'' treatment of the consistency factors gives fair but approximate results that tend to agree with the exact ones only at weak coupling. Since it is well known that the introduction of additional channels calls for the effective impurity coupling to be rescaled as $J/M$ \cite{Hewson,Cox&Z}, one can expect that the \textit{conventional compactification} becomes more reliable in the large-$M$ limit. However, the original claims were that compactification serves the purpose of (stabilizing and) moving the fixed point from intermediate to strong coupling. This could be in tension with the large-$M$ limit of the compactified model, and is also not observed in similarly compactified versions of the Anderson impurity model \cite{bulla1997}, (cf.~also the Majorana formulations of the two-channel Kondo model \cite{maldacena1997,*Zhang1999}). A renormalization-group analysis of these issues will be discussed in followup articles \cite{Ljepoja2024b,Ljepoja2024c}.
We shall find that, despite the differences highlighted here, the universal aspects of the low-energy physics do agree among the different formulations of the model, including the \textit{conventionally} compactified one, and the fixed points can be regarded as equivalent in that sense \footnote{It will be interesting to compare as well with results from integrability \cite{andrei1984,*andrei1995,*jerez1998,*bolech2002,*bolech2005a} and boundary conformal field theory \cite{affleck1991,*affleck1994,johannesson2003,*johannesson2005}.}.

The present work supports and complements our previous results for the Toulouse point of the two-lead Kondo model \cite{schiller1998a,bolech2016}, while further clarifying the role of the $n$-twiddle factors in the perturbative definition of that limit. Further work exploring the nonperturbative definition and algebraic properties of these unusual operators will be an important avenue for further studies. Speculatively, these would have the potential to open new field-theoretic paradigms for the subtle coupling of separate (dark) matter sectors to well established existing theories that were inferred using (low-order) perturbative calculations and matching experimental data (as in the celebrated Standard Model of particle physics \cite{Schwartz}).

\acknowledgements  
We acknowledge multiple discussions of our earlier work \cite{shah2016,bolech2016} that helped frame the present study. We are particularly grateful to N.~Andrei, H.U.~Baranger, D.~Feldman, M.P.A.~Fisher, T.~Giamarchi, A.W.W.~Ludwig and D.~Sen for their feedback. We also acknowledge the hospitality of the Kavli Institute for Theoretical Physics (KITP; supported by the NSF under Grant No.~PHY-1748958) and the International Centre for Theoretical Sciences (ICTS; Bangalore, India) where some of these discussions took place.

\appendix*

\section{More on Products of Vertex Operators} \label{Sec:Appendix}

One of the central ideas of the \textit{consistent} BdB framework is to point-split all products of fermion fields before the BdB manipulations. One then maintains the points separate during all the ensuing transformations and only takes the limit of coinciding points when ready to carry out the final debosonization step. 

The above procedure is what guarantees a consistent refermionization of the models (that has the corresponding implicit UV regularization) and avoids pitfalls when manipulating bosonized terms like, for instance, the spin-flip term $\psi^\dagger_{\uparrow\ell}(x)\psi_{\downarrow\ell}(x)$. Considering the vertex-product rendition of this term, not including Klein factors and multiplicative factors, and carrying out the usual manipulations, (cf.~with the ``conventional'' treatment in Ch.~28 Sec.~IV of Ref.~\onlinecite{Gogolin}, where point splitting is not done and the last step in the equation below would be an equality), one has
\begin{eqnarray}
    \lim_{x'\to x} e^{i\phi_{\uparrow\ell}(x)}e^{-i\phi_{\downarrow\ell}(x')} 
    &=& \lim_{x'\to x} e^{i[\phi_{\uparrow\ell}(x)-\phi_{\downarrow\ell}(x')]} \nonumber \\
    &\neq& e^{i[\phi_{s}(x)+\ell\phi_{sl}(x)]}
\end{eqnarray}
since, in the last step, although the contributions from $\partial\phi_{s}$ and $\partial\phi_{sl}$ can be dropped (because those fields are still present in the expression), those from $\partial\phi_{c}$ and $\partial\phi_{l}$
are those field's leading contributions and should not be ignored. As a result one has,
\begin{equation}
    \lim_{x'\to x} e^{i\phi_{\uparrow\ell}(x)}e^{-i\phi_{\downarrow\ell}(x')} 
    = \tilde{n}_{c} \tilde{n}_{l} \, e^{i[\phi_{s}(x)+\ell\phi_{sl}(x)]}
\end{equation}
where the $n$-twiddle factors are the notational tool we introduced to streamline the calculations and avoid these inconsistencies (see Eq.~(\ref{eq:bos})). Below, we summarize and extend some key \textit{consistent} BdB results presented previously in Sec.~IV of Ref.~\onlinecite{shah2016}.

During the \textit{point un-splitting} of the final debosonization step, one finds point-split products of so-called vertex operators of the type $\exp(i\lambda\phi)\exp(-i\lambda\phi)$ that have nontrivial operator product expansions (OPEs). The values of $\lambda$ depend on the particular bosonic transformations carried out; here we need to consider $\lambda=1$ and $\lambda=1/2$.

The first case relates directly to Mandelstam's formula (see Eq.~(\ref{Eq:Mandelstam})). The Klein factors drop out and one has
\begin{equation}
    \frac{e^{i\phi}}{\sqrt{2}}\frac{e^{-i\phi}}{\sqrt{2}}=\pi a \; \psi^{\dagger}\psi\to n
\end{equation}
The vertex operators on the left-hand side have been explicitly written with the normalization needed for their OPEs to be consistent with the anticommutator of lattice fermions. We can then, by convention, identify their product as a point-like fermionic-particle density $n$, (call it also $n^+$, and introduce the notation $n^-$ for the corresponding \textit{holes}). From the anticommutation properties (that derived in terms of the bosons require careful \textit{re-normal-ordering}, cf.~Ref.~\onlinecite{haldane1981}), it immediately follows that $n^2=n$ and $n^+n^-=0$, which we refer to as \textit{idempotence} and \textit{co-nilpotence}, respectively. One also has $n^++n^-=1$.

The $\lambda=1/2$ case is closely related but even more delicate. This is what prompts the definition of the $n$-twiddle factors, 
\begin{equation}
\label{eq:a2}
    \tilde{n}=\frac{1}{\sqrt{2}}e^{i\phi/2}e^{-i\phi/2}\to \sqrt{n}
\end{equation}
where the last identification is done by comparing OPEs (and other careful considerations that we do not repeat in here), and double checked by a composite-operator OPE showing that $\tilde{n}^2\to n$. The property of co-nilpotence is directly inherited from Eq.~(\ref{eq:a2}), since $\tilde{n}^+\tilde{n}^-\to\sqrt{n^+n^-}=0$. 
The fact that one is using the positive branch of the square root function, together with the positive definiteness of the sum $\tilde{n}^++\tilde{n}^-$, gives the ``resolution of the identity'' for the $n$-twiddles, $\tilde{n}^++\tilde{n}^-\to1$, since
\begin{equation}
    (\tilde{n}^++\tilde{n}^-)^2=(\tilde{n}^+)^2+(\tilde{n}^-)^2\to n^++n^-=1
\end{equation}

Just like co-nilpotence, the property of idempotence ($\tilde{n}^2=\tilde{n}$) is also natural but needs to be adopted separately, and the considerations given in the main text (see Sec.~\ref{sec:flat-band}, as well as the subsequent renormalization-group studies \cite{Ljepoja2024b,Ljepoja2024c}), show that it needs to be supplemented by the additional property of \textit{independent n-twiddle histories} for high-order diagrammatic processes. This is because the $n$-twiddle factors entering the refermionized interaction terms correspond to different-sectors ($c$ and $l$) than the individual fermions present in those terms ($s$ and $sl$) and therefore have quantum and thermal fluctuations independent of each other and of the other sectors. 
Taken together, these three properties guarantee the matching of all results of the original model and the one refermionized using consistent BdB (which does not happen when conventional BdB is used instead).

\input{Paper1BdB.bbl}

\end{document}

%% file: Paper1BdB.bbl
%